\documentclass{article}

\usepackage{amsmath,amsfonts,bm}

\def\eqref#1{equation~\ref{#1}}

\def\1{\bm{1}}

\DeclareMathAlphabet{\mathsfit}{\encodingdefault}{\sfdefault}{m}{sl}
\SetMathAlphabet{\mathsfit}{bold}{\encodingdefault}{\sfdefault}{bx}{n}

\usepackage{amsmath, amssymb, mathtools}
\usepackage{amsthm}
\usepackage{amsfonts}
\usepackage{nicefrac}

\usepackage[utf8]{inputenc}
\usepackage[T1]{fontenc}
\usepackage{microtype}

\usepackage[accepted]{icml2025}

\usepackage{xurl}       
\usepackage{url}

\usepackage[hidelinks]{hyperref}

\usepackage[capitalize,noabbrev]{cleveref}

\usepackage{graphicx}
\usepackage{subcaption}
\usepackage{float}
\usepackage{wrapfig}
\usepackage{tikz}
\usepackage{diagbox}

\usepackage{booktabs}
\usepackage{tabularx}
\usepackage{colortbl}
\usepackage{multirow}
\usepackage{makecell}

\usepackage{xcolor}
\usepackage{tcolorbox}

\usepackage{listings}
\usepackage[inline]{enumitem}
\usepackage{soul}            
\usepackage[normalem]{ulem}  

\definecolor{lightgreen}{rgb}{0.56, 0.93, 0.56}
\definecolor{lightcoral}{rgb}{0.94, 0.50, 0.50}
\definecolor{chatgptcolor}{HTML}{1f77b4}
\definecolor{llamacolor}{HTML}{ff7f0e}
\definecolor{highlightgreen}{rgb}{0.56, 0.93, 0.56}
\definecolor{highlightred}{rgb}{0.94, 0.50, 0.50}
\definecolor{pastelgreen}{RGB}{197, 239, 197}
\definecolor{pastelred}{RGB}{230, 184, 184}

\newcommand{\whitecircle}{\tikz[baseline=-0.5ex]\draw (0,0) circle (0.8ex);}
\newcommand{\whitesquare}{\tikz[baseline=0.2ex]\draw (0,0) rectangle (1.4ex,1.4ex);}

\newcommand{\rbox}[1]{\sethlcolor{pastelred}\hl{#1}}
\newcommand{\gbox}[1]{\sethlcolor{pastelgreen}\hl{#1}}

\icmltitlerunning{Optimizing Adaptive Attacks against Watermarks for Language Models }

\begin{document}

\twocolumn[
\icmltitle{Optimizing Adaptive Attacks against Watermarks for Language Models  }

\begin{icmlauthorlist}
\icmlauthor{Abdulrahman Diaa}{uw}
\icmlauthor{Toluwani Aremu}{mbzuai}
\icmlauthor{Nils Lukas}{mbzuai}
\end{icmlauthorlist}

\icmlaffiliation{uw}{David R. Cheriton School of Computer Science, University of Waterloo, Ontario, Canada}
\icmlaffiliation{mbzuai}{Mohammed Bin Zayed University of Artificial Intelligence (MBZUAI), Abu Dhabi, UAE}

\icmlcorrespondingauthor{Abdulrahman Diaa}{abdulrahman.diaa@uwaterloo.ca}
\icmlcorrespondingauthor{Nils Lukas}{nils.lukas@mbzuai.ac.ae}

\icmlkeywords{watermarking, large language models, adaptive attacks, robustness, paraphrasing, reinforcement learning, text generation}

\vskip 0.3in
]



\printAffiliationsAndNotice{}  

\begin{abstract}
 Large Language Models (LLMs) can be misused to spread unwanted content at scale. Content watermarking deters misuse by hiding messages in content, enabling its detection using a secret \emph{watermarking key}. Robustness is a core security property, stating that evading detection requires (significant) degradation of the content's quality. Many LLM watermarking methods have been proposed, but robustness is tested only against \emph{non-adaptive} attackers who lack knowledge of the watermarking method and can find only suboptimal attacks. We formulate watermark robustness as an objective function and use preference-based optimization to tune \emph{adaptive} attacks against the specific watermarking method. Our evaluation shows that (i) adaptive attacks evade detection against all surveyed watermarks, (ii) training against \emph{any} watermark succeeds in evading unseen watermarks, and (iii) optimization-based attacks are cost-effective. Our findings underscore the need to test robustness against adaptively tuned attacks. We release our adaptively tuned paraphrasers at \url{https://github.com/nilslukas/ada-wm-evasion}.
\end{abstract}
\section{Introduction}
\begin{figure}
    \centering
    \includegraphics[width=\linewidth]{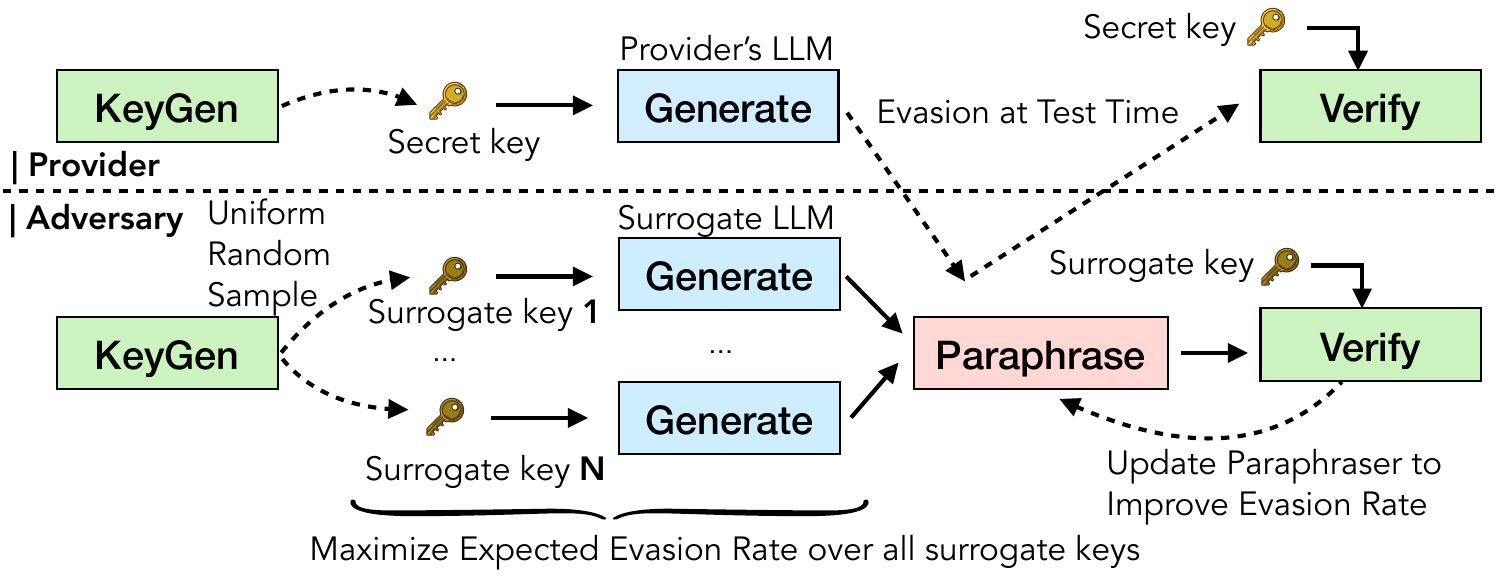}
    \caption{Adaptive attackers know the watermarking algorithms (\textsc{KeyGen}, \textsc{Verify}), but not the secret key, so they can optimize a paraphraser against a specific watermark. }
    \label{fig:diagram}
\end{figure}
A few Large Language Model (LLM) providers empower many users to generate human-quality text at scale, raising concerns about dual use~\citep{barrett2023identifying}.
Untrustworthy users can \emph{misuse} the provided LLMs to generate harmful content, such as online spam~\citep{weidinger2021ethical}, misinformation~\citep{chen2024combatingmisinformation}, or to facilitate phishing attacks~\citep{shoaib2023deepfakes}.
The ability to detect generated text can control these  risks~\citep{grinbaum2022ethical}.

Content watermarking enables the detection of generated outputs by embedding hidden messages that can be extracted with a secret watermarking key.
Some LLM providers, such as \citet{deepmind_synthid} and Meta~\citep{sanroman2024proactive}, have already deployed watermarking to promote the ethical use of their models.
A threat to these providers are users who perturb generated text to evade watermark detection while preserving text quality. 
Such undetectable, generated text could further erode trust in the authenticity of digital media~\citep{aiexecutiveorder}.  

A core security property of watermarking is \emph{robustness}, which requires that evading detection is only possible by significantly degrading text quality.
Testing robustness requires identifying the most effective attack against a specific watermarking method.
However, existing content watermarks for LLMs~\citep{kirchenbauer2023watermark, aaronson2023watermarking, christ2023undetectable, kuditipudi2023robust} test robustness only against \emph{non-adaptive} attackers, who lack knowledge of the watermarking algorithms.  
This reliance on obscurity makes watermarking vulnerable to \emph{adaptive} attacks~\citep{lukas2024leveraging, jovanovic2024watermark} when information about the watermarking algorithms is leaked. 

We propose a method to curate preference datasets and adaptively optimize an attack against \emph{known} content watermarking algorithms. 
Optimization is challenging due to (i) the complexity of optimizing within the discrete textual domain and (ii) the limited computational resources available to attackers.
We demonstrate that adaptively tuned, open-weight LLMs such as $\texttt{Llama2-7b}$~\citep{touvron2023llama} evade detection at negligible impact on text quality against $\texttt{Llama3.1-70b}$~\citep{dubey2024llama}. 
Our attacker spends less than 7 GPU hours to achieve an evasion rate of over $96\%$ against any of the surveyed watermarking methods with negligible impact on text quality. 
Our attacks are Pareto optimal, even in the non-adaptive setting where they must transfer to unseen watermarks.
Hence, future watermarking methods must consider our attacks to test robustness.
%

We make the following contributions.
\begin{enumerate*}[label=(\arabic*), before={\unskip{ }}, itemjoin={{ }}, itemjoin*={{ }}]
    \item We propose methods to curate preference-based datasets using LLMs, enabling us to adaptively fine-tune watermark evasion attacks against state-of-the-art language watermarks.
    \item Adaptively tuned paraphrasers with 0.5-7 billion parameters evade detection from all tested watermarks at a negligible impact on text quality. 
    We demonstrate their Pareto optimality for evasion rates greater than 90\%\footnote{{Closed models such as GPT-4o are also on the Pareto front (due to high text quality) but achieve lower evasion rates.}}.
    Optimization against models with $46\times$ more parameters requires less than seven GPU hours, which challenges security assumptions, as even adversaries with limited resources can reliably evade detection using our attacks. 
    \item We test our attacks in the non-adaptive setting against unseen watermarks and demonstrate that they {remain Pareto optimal} compared to other {non-adaptive} attacks.
    Our results underscore the necessity of using optimizable, adaptive attacks to test robustness. 
    \item {We publicly release our adaptively tuned paraphrasers to facilitate further research on robustness against adaptive attackers.}
\end{enumerate*}

\section{Background}
\label{sec:background}

\textbf{Large Language Models (LLMs)} estimate the probability distribution of the next token over a vocabulary $\mathcal{V}$ given a sequence of tokens. 
Autoregressive LLMs predict each subsequent token based on all preceding tokens. 
Formally, for a sequence of tokens $x_1, \ldots, x_n$, an LLM models:
\[P(x_n | x_1, \ldots, x_{n-1}) = \text{softmax}(f_\theta(x_1, \ldots, x_{n-1}))_n\]

where $f_\theta$ is a neural network with parameters $\theta$.
Optimizing LLMs to maximize a reward function is challenging because the text is discrete, and the autoregressive generation process prevents direct backpropagation through the token sampling steps~\citep{schulman2017proximal}.

\textbf{LLM Content Watermarking} hides a message in generated content that can later be extracted with access to the content using a secret watermarking key. 
A \emph{watermarking method}, as formalized by~\citep{lukas2024leveraging}, comprises a set of algorithms (\textsc{KeyGen}, \textsc{Embed}, \textsc{Verify}):
\begin{itemize}
    \item $\tau \gets \textsc{KeyGen}(\theta,\gamma)$: A randomized function to generate a watermarking key $\tau$ given secret (i) LLM parameters $\theta$ and (ii) random seeds $\gamma \in \mathbb{R}$. 
    \item $\theta^* \gets \textsc{Embed}(\theta, \tau, m)$: Given a LLM $\theta$, a watermarking key $\tau$ and a message $m$, this function\footnote{\textsc{Embed} can modify the entire inference process.} returns parameters $\theta^*$ of a \emph{watermarked} LLM that generates watermarked text.
    \item $\eta \gets \textsc{Verify}(x, \tau, m)$: Detection involves (i) extracting a message $m'$ from text $x$ using $\tau$ and (ii) calculating the $p$-value $\eta$ for rejecting the null hypothesis that $m$ and $m'$ match by chance. 
\end{itemize}
\textbf{$(\epsilon, \delta)$-Robustness.} A text watermark is a hidden signal in text that can be mapped to a message $m\in \mathcal{M}$ using a secret watermarking key $\tau$.
The key $\tau$ refers to secret random bits of information used for detecting a watermark.
A watermark is \emph{retained} if \textsc{Verify} outputs $\eta< \rho$, for $\rho \in \mathbb{R}^+$. 
Let $Q: \mathcal{V}^*\times \mathcal{V}^* \rightarrow \mathbb{R}$ be a function to measure text quality between pairs of texts.
We say that a watermark is $(\epsilon, \delta)$-robust if any paraphrase $y = \mathcal{A}(x)$ of a watermarked text $x$ that remains high-quality (i.e., $Q(x,y)>\delta$) also retains the watermark with probability $\ge 1-\epsilon$. 
Let $\mathcal{A}$ be a randomized paraphrasing method, then robustness can be stated as follows. 
\begin{align}
    \label{eq:robustness}
    \underset{y\gets \mathcal{A}(x)}{\text{Pr}}\left[ \textsc{Verify}(y, \tau, m) \geq \rho ~\land~Q(x,y) > \delta  \right] < \epsilon
\end{align} 
\textbf{Evasion Attacks.}
Watermark evasion attacks are categorized by the attacker's access to the provider's (i) LLM, (ii) detection algorithm \textsc{Verify} that uses the provider's secret watermarking key, and (iii) knowledge of the watermarking algorithms. 
{A \emph{no-box} attacker has no access to the provider's LLM}, whereas \emph{black-box} attackers have API access, and \emph{white-box} attackers know the parameters of the provider's LLM.
\emph{Online} attackers can query the provider's \textsc{Verify} functionality, as opposed to \emph{offline} attackers who have no such access. 
\emph{Adaptive} attackers know the algorithmic descriptions (\textsc{KeyGen}, \textsc{Embed}, \textsc{Verify}) of the provider's watermarking method, while \emph{non-adaptive} attackers lack this knowledge.
Our work focuses on {no-box}, offline attacks in adaptive and non-adaptive settings. 

\textbf{Surveyed Watermarking Methods.}
Following \cite{piet2023mark}, we evaluate the robustness of four state-of-the-art watermarking methods\footnote{In \Cref{appendix:baseline_tests}, we evaluate against more watermarks including SynthID~\citep{dathathri2024scalable}, Unigram~\citep{zhao2024provable} and SIR~\citep{liu2024a}.}
The \texttt{Exp}~\citep{aaronson2023watermarking} method marks text by selecting tokens that maximize a score combining the conditional probability $P(x_n \mid x_0 \dots x_{n-1})$ and a pseudorandom value derived from a sliding window of prior tokens.
The \texttt{Dist-Shift}~\citep{kirchenbauer2023watermark} method favours tokens from a green list, which is generated based on pseudorandom values and biases their logits to increase their selection probability.
The \texttt{Binary}~\citep{christ2023undetectable} approach converts tokens into bit-strings determined by pseudorandom values and the language model's bit distribution, subsequently translating the bit-string back into a token sequence. 
Lastly, the \texttt{Inverse}~\citep{kuditipudi2023robust} scheme uses inverse transform sampling by computing a cumulative distribution function ordered pseudorandomly according to a secret key and using a fixed pseudorandom value to sample from this distribution.
We refer to \cite{piet2023mark} for more details.

\section{Threat Model}
We consider a provider capable of training LLMs and deploying them to many users via a black-box API, such as Google with Gemini or OpenAI with ChatGPT. 
The threat to the provider are untrustworthy users who misuse the provided LLM and generate harmful content without detection.

\textbf{Provider's Capabilities and Goals} \textit{(Deployment)} The provider fully controls the LLM and its text generation process, including the ability to embed a watermark into generated text. 
\textit{(Watermark Verification)} The provider must be able to verify their content watermark in each generated text. 
Their goal is to have a watermark that is (i) quality-preserving and (ii) robust, enabling detection of generated text at a given, low False Positive Rate (FPR) $\rho\in \mathbb{R}^+$. 

\textbf{Attacker's Capabilities.} 
\textit{(Access Restrictions)} We consider a (i) {no-box attacker who cannot collect any watermarked texts during training} and is (ii) offline, meaning that they cannot access the provider's \textsc{Verify} function.
Our focus is on (iii) adaptive attackers, who know the provider's watermark algorithms (\textsc{KeyGen}, \textsc{Embed}, \textsc{Verify}) but do not know the secret inputs used for watermarking, such as random seeds or the provider's LLM.
We also evaluate how adaptive attacks transfer in the non-adaptive setting against unseen watermarks. 
\textit{(Surrogate Models)} A surrogate model is a model trained for the same task as the provider's model. 
For example, while GPT-4o's weights are not public, the attacker can access the parameters of smaller, publicly available models such as those from the \texttt{Llama2}~\citep{touvron2023llama} model family. 
Our attacker can use such open-weight \emph{surrogate} models for \emph{paraphrasing} text.
We assume the surrogate model's text quality is inferior to the provided model; otherwise, there would be no need to use the watermarked model. 
\textit{(Compute)} Our attacker has limited computational resources and cannot train LLMs from scratch.

\textbf{Attacker's Goals.} The attacker wants to use the provided, watermarked LLM to generate text (i) without a watermark and (ii) with high quality. 
We measure text quality using many metrics, including a quality function $Q: \mathcal{V}^* \times \mathcal{V}^* \rightarrow \mathbb{R}$ between pairs of text when the attacker attempts to evade detection. 
We require that the provider correctly verifies their watermark with a given p-value threshold at most $\rho$. 
Lower thresholds make evasion more likely to succeed, i.e., detection becomes more challenging for the provider.

\textbf{Our motivation} is to evaluate the robustness of watermarking against constrained attackers that (i) have limited resources and (ii) lack any information about the watermarking key and samples. If successful attacks exist in this pessimistic no-box setting, the provider cannot hope to have a robust watermark against more capable attackers (e.g., with black-box access). We show that (i) such attacks exist, (ii) they are inexpensive, and (iii) they do not require access to watermarked samples. We believe the development of defenses should focus on the no-box setting first.

\section{Related Work}
\label{sec:related-work}
We evaluate the robustness of \emph{content} watermarking~\citep{lukas2023ptw} methods against no-box, offline attackers in the adaptive and non-adaptive settings (see \Cref{sec:background}).
%
%
Other watermark evasion attacks, including those by \citet{hu2024stable}, \citet{kassis2024unmarker}, and \citet{lukas2024leveraging}, focus on the image domain, whereas our work focuses on LLMs. 
{\citet{jovanovic2024watermark,pang2024no} propose black-box attacks against LLMs that require collecting many watermarked samples under the same key-message pair. We focus on no-box attacks.}
%
\citet{jiang2023evading} propose online attacks with access to the provider's watermark verification, whereas we focus on a less capable \emph{offline} attacker who cannot verify the presence of the provider's watermark. 
Current attacks are either non-adaptive, such as DIPPER~\citep{krishna2023paraphrasing} or handcrafted against one watermark~\citep{jovanovic2024watermark}.
We focus on optimizable, adaptive attacks and show that they remain effective in the non-adaptive setting.

%
\citet{zhang2024watermarks} demonstrated the impossibility of robust watermarking against attackers with access to quality and perturbation oracles, showing that random walks with the perturbation oracle provably removes watermarks. 
Our approach differs in that it adaptively optimizes to find a single-step perturbation for evading watermark detection. 
We demonstrate the feasibility and efficiency of our attacks, achieving watermark evasion at low computational cost (USD $\leq 10\$$).

\section{Conceptual Approach}
Our goal is to adaptively fine-tune an open-weight paraphraser $\theta_P$ against known watermarking methods. 
The attacker lacks knowledge of the provider's watermarking key $\tau \leftarrow \textsc{KeyGen}(\theta, \gamma)$, which depends on (i) the unknown random seed $\gamma$ and (ii) the unknown parameters $\theta$ of the provider's LLM. 
Our attacker overcomes this uncertainty by choosing an open-weight surrogate model $\theta_S$ to generate so-called \emph{surrogate} watermarking keys $\tau'$ and optimizes the expected evasion rate over many random seeds $\gamma\sim \mathbb{R}$.

\subsection{Robustness as an Objective Function} Let $P_\theta: \mathcal{V}^* \rightarrow \mathcal{V}^*$ denote a randomized paraphrasing function\footnote{We consider language models as paraphrasers, where randomness arises from sampling the next token.}, $H_\theta: \mathcal{V}^* \rightarrow \mathcal{V}^*$ is a function to generate text given a query $q\in \mathcal{T} \subseteq \mathcal{V}^*$ and $Q: \mathcal{V}^* \times \mathcal{V}^* \rightarrow \mathbb{R}$ measures the similarity between pairs of text. 
We formulate robustness using the objective function in \Cref{eq:attack_objective} that optimizes the parameters $\theta_P$ of a paraphrasing model. 


\begin{equation}
  \label{eq:attack_objective}
  \begin{aligned}
    \max_{\theta_P}\quad
    &\mathbb{E}_{\substack{
        \gamma \sim \mathcal{R}\\
        m' \sim \mathcal{M}\\
        q \sim \mathcal{T}
    }}
    \Bigl[
    \mathbb{E}_{\substack{
        \tau' \gets \textsc{KeyGen}(\theta_S, \gamma)\\
        \theta_S^* \gets \textsc{Embed}(\theta_S, \tau', m')\\
        x \gets H(\theta_S^*, q)\\
        x' \gets P(\theta_P, x)
    }}
    \\[1ex]
    &\quad
      \textsc{Verify}\bigl(x',\,\tau',\,m'\bigr)
      \;+\;
      Q\bigl(x',\,x\bigr)
    \Bigr].
  \end{aligned}
\end{equation}

\Cref{eq:attack_objective} finds optimal parameters for the paraphraser $\theta_P$ by sampling uniformly at random over (i) random seeds $\gamma \sim \mathbb{R}$, (ii) messages $m' \sim \mathcal{M}$ and (iii) queries $q\sim \mathcal{T}$.
The second expectation is taken over a \emph{surrogate watermarking key}, generated using knowledge of the $\textsc{KeyGen}$ algorithm, the surrogate model's parameters $\theta_S$ and a (previously sampled) random seed $\gamma$ as input. 
The surrogate model, key, and message are used to embed a watermark into the surrogate model $\theta_S^*$ (with knowledge of $\textsc{Embed}$), which  generates a watermarked sample $x$. 
The optimization process finds optimal parameters $\theta_P^*$ such that the paraphraser has a high probability of generating text $y\gets P(\theta_P, x)$ that evades watermark detection and preserves text quality compared to $x$. 
Note that knowledge of the watermarking algorithms (\textsc{KeyGen}, \textsc{Embed}, \textsc{Verify}) is required to generate surrogate keys needed to optimize \Cref{eq:attack_objective}.

Optimization presents multiple challenges.
The attacker optimizes over different random seeds $\gamma$ and a surrogate model $\theta_S$ than those used by the provider, since our attacker does not know the provider's model parameters $\theta$ or random seeds. 
This lack of knowledge adds uncertainty for the attacker. 
The discrete nature of text and the inability to backpropagate through its generation process make maximizing the reward challenging~\cite{shin2020autoprompt}.
Furthermore, the reward function depends on $\textsc{Verify}$, which may not be differentiable. 
Deep reinforcement learning (RL) methods~\citep{schulman2017proximal, rafailov2024direct} do not require differentiable reward functions. 
However, RL is known to be compute-intensive and unstable, making it unclear whether optimization can achieve a high reward using limited computational resources. 
    \begin{algorithm}[H]
    \caption{curates a preference dataset to optimize the adaptive attack's objective in \cref{eq:attack_objective}.}
    \label{alg:preference-dataset-collection}
    \begin{algorithmic}[1]
    \REQUIRE Surrogate $\theta_S$, Paraphraser $\theta_P$, Queries $\mathcal{T}$, Messages $\mathcal{M}$, Paraphrase Repetition Rate $N$, False Positive Rate Threshold $\rho$, Quality Threshold $\delta$
    \STATE $\mathcal{D} \gets \emptyset$ // \textit{The preference dataset}
    \STATE // \textit{Sample from known watermarking methods $\mathcal{W}$}
    \FOR{$(\textsc{KeyGen}, \textsc{Embed}, \textsc{Verify}) \in \mathcal{W}$}
        \FOR{each $q \in \mathcal{T}$}
            \STATE $m \sim \mathcal{M}$
            \STATE $\tau' \gets \textsc{KeyGen}(\theta_S, \textsc{Rnd}())$
            \STATE $\theta^*_S \gets \textsc{Embed}(\theta_S, \tau', m)$
            \STATE $r \gets S_{\theta^*_S}(q)$ // \textit{Watermarked text under $\tau$'}
            \STATE // \textit{If watermark can be detected}
            \IF{$\textsc{Verify}(r, \tau', m) < \rho$}
                \STATE // \textit{Rejected (0) and Chosen (1) paraphrases}
                \STATE $R^0, R^1 \gets \emptyset, \emptyset$
                \FOR{$i \in [N]$}
                    \STATE $r' \gets P_{\theta_P}(r)$ // \textit{Paraphrase (randomized)}
                    \STATE $a \gets \mathbf{1}{[Q(r, r') \geq \delta]}$
                    \STATE $b \gets a \cdot \mathbf{1}[\textsc{Verify}(r', \tau', m) > \rho]$
                    \STATE $R^b \gets R^b \cup \{r'\}$
                \ENDFOR
                \FOR{$j \in [|R^1|]$} 
                    \STATE $r_n' \gets (j \leq |R^0|) \; ? \; R^0_j : r$ 
                    \STATE $\mathcal{D} \gets \mathcal{D} \cup \{(r, r_n', R^1_j)\}$ // \textit{Match pairwise}
                \ENDFOR
            \ENDIF
        \ENDFOR
    \ENDFOR
    \RETURN $\mathcal{D}$ // \textit{The preference dataset}
    \end{algorithmic}
\end{algorithm}
\subsection{{Preference Dataset Curation}}
We use reinforcement learning (RL) methods such as Direct Preference Optimization (DPO)~\citep{rafailov2024direct} to optimize \Cref{eq:attack_objective}.
However, DPO requires collecting a preference dataset of \emph{positive} and \emph{negative} examples to fine-tune the paraphraser.
A \emph{negative} sample is one that retains the watermark, representing a failed attempt at watermark evasion. 
In contrast, positive samples do not retain a watermark and have a high text quality $Q(r,r_p')>\delta$ for an attacker-chosen $\delta \in \mathbb{R}^+$. 
To bootstrap optimization, we require the ability to curate positive and negative examples, which we achieve by using a publicly available, open-weight paraphrasers such as Llama2-7b. 
We curate triplets $(r,r_n',r_p')$ via best-of-N rejection sampling. These triplets contain a watermarked sample $r$ and two paraphrased versions, $r_n'$ and $r_p'$, representing the negative and positive examples, respectively. 
\Cref{alg:preference-dataset-collection} implements the algorithm to curate our preference dataset.

\Cref{alg:preference-dataset-collection} randomly samples from a set of known watermarking methods $\mathcal{W}$ (line 3) and from the set of task-specific queries $\mathcal{T}$ (line 4).
It samples a message $m$ (line 5) and generates a surrogate watermarking key $\tau'$ to embed a watermark into the surrogate generator (lines 6-7). 
We generate text $r$ using the watermarked model $\theta^*_S$ (line 8) and verify whether it retains the watermark (line 9).
The paraphrase model $\theta_P$ generates $N$ paraphrased versions of $r$ that we partition into positive and negative samples (lines 13-17).
A sample $r_p'$ is positive ($b=1$) if it does not retain the watermark and has high text quality ($\geq \delta$); otherwise, it is negative ($r_n'$, $b=0$). 
%
%
For each positive sample, we select one corresponding negative sample and add the watermarked text and the negative and positive paraphrases to the preference dataset $\mathcal{D}$ (lines 19-21). 

\begin{table*}[ht]
    \centering
    \small
    \begin{tabularx}{\linewidth}{lX} 
        \toprule
        Attack Name & Description \\
        \cmidrule(lr){1-2}
        DIPPER~\citep{krishna2023paraphrasing} & Train an 11b Sequence-to-Sequence model for paraphrasing.  \\
        Translate~\citep{piet2023mark} & Translate to another language and back (e.g., French, Russian). \\ 
        Swap~\citep{piet2023mark} & Randomly remove, add or swap words. \\
        Synonym~\citep{piet2023mark} & Replace words with a synonym using WordNet~\citep{miller1995wordnet}. \\
        HELM~\citep{bommasani2023holistic} & Randomly add typos, lowercase or contractions. \\ 
        Llama, Qwen2.5, GPT3.5  & Paraphrase text using a publicly accessible LLM.  \\ 
        \midrule
        Ours-\texttt{Llama2-7b}-\texttt{Exp} & Paraphrase with a \texttt{Llama2-7b} model tuned adaptively against \texttt{Exp}.\\
        \bottomrule
    \end{tabularx}
    \caption{\label{tab:evasion-attack}(Top) The non-adaptive baseline attacks we consider in our study against (Bottom) our adaptively fine-tuned attacks.
    We refer to \cite{piet2023mark} for details on the baseline attacks and \Cref{appendix:attack-descriptions} for our adaptive attack.}
\end{table*}

\section{Evaluation}
\label{sec:experiments}
We report all runtimes on NVIDIA A100 GPUs accelerated using VLLM~\citep{kwon2023efficient} for inference and DeepSpeed~\citep{deepspeed} for training.
Our implementation uses PyTorch and the Transformer Reinforcement Learning (TRL) library~\citep{vonwerra2022trl}. We use the open-source repository by \citet{piet2023mark}, which implements the four surveyed watermarking methods.
We test robustness using hyper-parameters suggested by \cite{piet2023mark}.
Please refer to \Cref{appendix:watermark-configurations} for details on hyperparameter selection and generalizability of our attacks against a range of hyperparameters.
All LLMs used in our evaluations have been instruction-tuned.   
A detailed description of our attack setup, including prompting strategies and training hyperparameters, is available in \Cref{appendix:attack-descriptions}.
\Cref{tab:evasion-attack} summarizes other surveyed evasion attacks.

\subsection{Preference Dataset Collection}
\begin{figure}
    \centering
    \includegraphics[width=\linewidth]{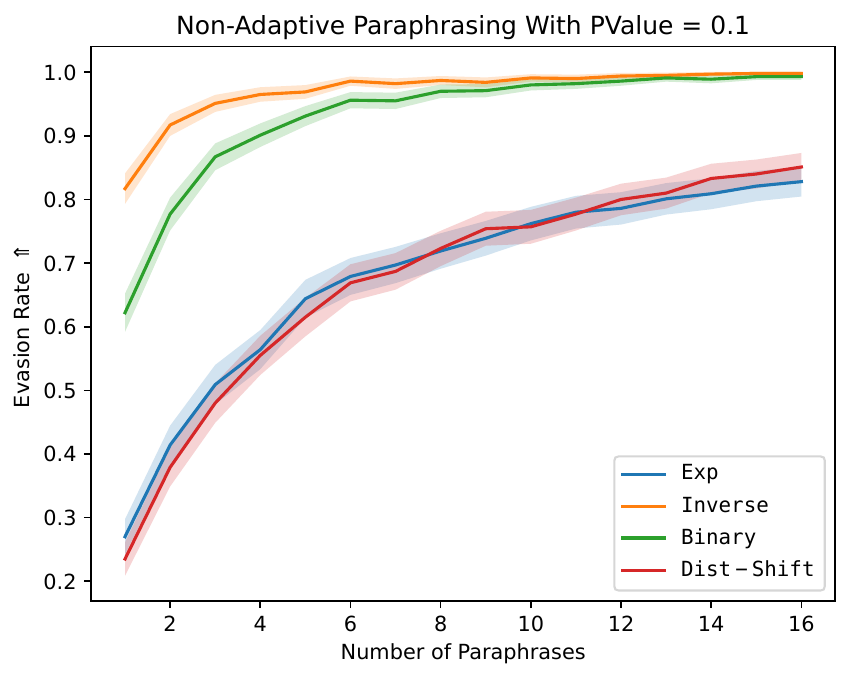}
    \caption{\Cref{alg:preference-dataset-collection} paraphrases text $N$ times in lines 13-17. This graph shows the expected evasion rate of the best sample (lines 15-17) for the number of paraphrases using a vanilla \texttt{Llama2-7b} as the paraphraser.}
    \label{fig:non-adaptive-paraphrasing-repetition}
\end{figure}

For a given watermarked sequence generated by the surrogate model, the attacker generates $N$ paraphrased versions using the non-optimized paraphraser and calculates the {best-of-N evasion rate} with the surrogate key (\Cref{alg:preference-dataset-collection}, lines 9-12). 
\Cref{fig:non-adaptive-paraphrasing-repetition} shows the number of repetitions $c$ needed to achieve a given evasion rate across four watermarking methods using \texttt{Llama2-7b} as both the surrogate and paraphrasing model. 
Our attacker can choose the best-of-N paraphrases because they know the surrogate watermarking key to detect a watermark.
The attacker cannot choose the best-of-N paraphrases against the provider's watermarked text, as they lack access to the provider's key. 

\Cref{fig:non-adaptive-paraphrasing-repetition} shows the success rate of observing at least one positive sample after N paraphrases against methods designed for robustness (\texttt{Dist-Shift}, \texttt{Exp}) and undetectability (\texttt{Inverse}, \texttt{Binary}).
{The attacker requires limited computational resources to curate a large preference dataset against any of the four surveyed watermarks.}
For instance, to collect $|D|=7\,000$ 
preference samples, each of $T=512$ tokens, at a rate of $1\,800$ tokens/second, we expect this to take approximately 1.5 GPU hours for \texttt{Dist-Shift}, but only 0.5 GPU hours for \texttt{Inverse}. 
In practice, including the overhead of evaluating quality and detecting watermarks, we require less than 5 GPU hours to curate $7\,000$ samples for \texttt{Dist-Shift}.
At current AWS rates, an attacker who uses our attacks faces only negligible costs of less than \$10 USD to curate a preference dataset containing $7\,000$ samples and fine-tune the paraphraser. 
Further details on the curation of the prompt sets used for training and evaluation are provided in~\Cref{appendix:prompt-set-curation}.

\subsection{Ablation Studies}

In our experiments, we ablate over the following settings. 

\begin{enumerate*}[label=(\arabic*), before={\unskip{ }}, itemjoin={{, }}, itemjoin*={{, and }}]
    \item \textbf{Adaptivity}: (\emph{Adaptive}) The same watermarking method is used for training and testing. 
    (\emph{Non-adaptive}) The attack is tested against unseen watermarking methods.
    \item \textbf{Target Models}: We evaluate 2 models used by the provider: \texttt{Llama2-13b}, \texttt{Llama3.1-70b}.
    \item \textbf{Attacker's Models}: Our attacker matches surrogate and paraphrasing models. We consider \texttt{Llama2}~\citep{touvron2023llama} and \texttt{Qwen2.5}~\citep{qwen2.5} from $0.5$b to $7$b parameters.
    \item \textbf{Watermarking Methods}: \texttt{Exp}~\citep{aaronson2023watermarking}, \texttt{Dist-Shift}~\citep{kirchenbauer2023reliability}, \texttt{Inverse}~\citep{kuditipudi2023robust}, \texttt{Binary}~\citep{christ2023undetectable}. 
    \item \textbf{Hyper-Parameters}: We ablate over multiple hyper-parameters that a provider can choose (see  \Cref{appendix:watermark-configurations}).
    \item \textbf{False Positive Rates (FPR)}:  \Cref{appendix:fpr_ablation} ablates over $\rho \in \{0.01, 0.025,0.05,0.075,0.1\}$ when the provider can tolerate higher FPR thresholds for detection.
\end{enumerate*}

A watermark has been \emph{retained} if the null hypothesis that the watermark is not present in the content can be rejected with a given p-value specified by the provider. 
The \emph{evasion rate} is calculated as the fraction of watermarked text that does not retain the watermark after applying the paraphrasing attack. 
Due to the lack of a gold-standard metric to assess text quality, we measure quality with multiple metrics: LLM-Judge, LLM-CoT, and LLM-Compare from \citet{piet2023mark}, Mauve~\citep{pillutla2021mauvemeasuringgapneural}, and Perplexity (PPL) with \texttt{Llama3-8B-Instruct}.
{To enhance clarity, we only report the LLM-Judge metric in the main paper following \citet{piet2023mark}.} Full descriptions of all quality metrics are provided in \Cref{appendix:metrics}.
Unless otherwise specified, we use a p-value threshold of $\rho=0.01$.

\subsection{Experimental Results}
\label{sec:experimental_results}

\begin{figure*}[t] 
    \centering
    \begin{subfigure}{0.49\linewidth} 
        \centering
        \includegraphics[height=4.9cm]{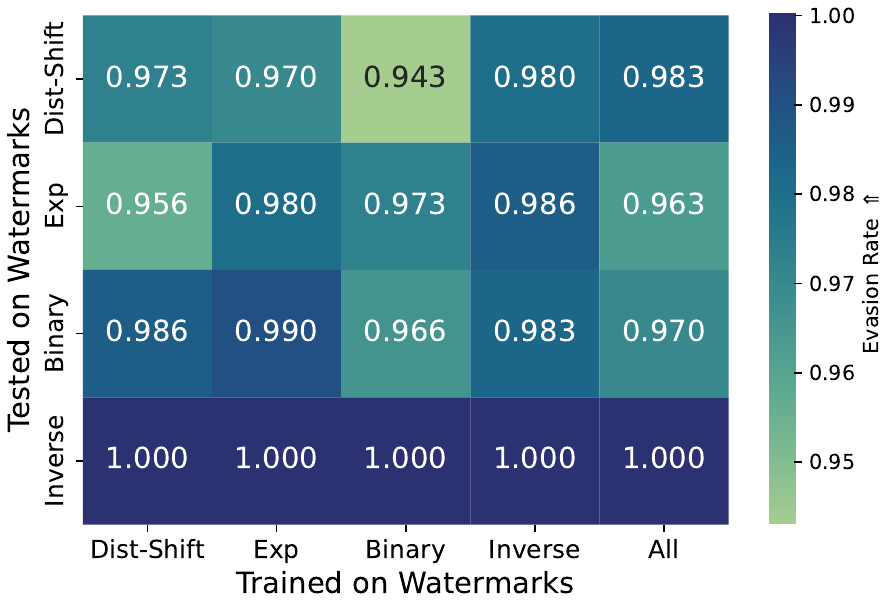}
        \label{fig:sub1}
    \end{subfigure}
    \hfill
    \begin{subfigure}{0.49\linewidth} 
        \centering
        \includegraphics[height=4.9cm]{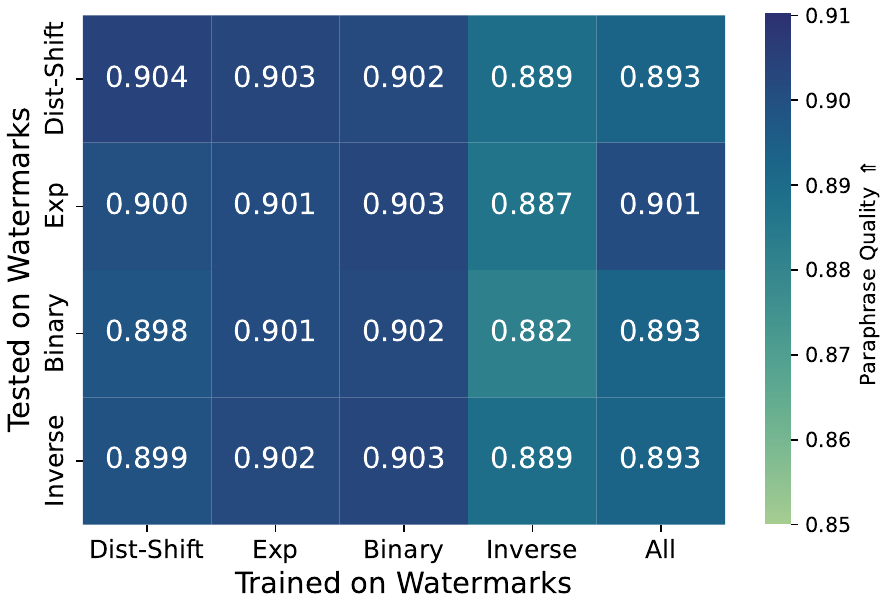}
        \label{fig:sub2}
    \end{subfigure}
    \caption{The evasion rates (Left) and text quality measured with LLM-Judge (Right). The attacker uses a matching \texttt{Llama2-7b} surrogate and paraphraser model versus the provider's \texttt{Llama2-13b}. Results for adaptive attacks are on the diagonal. For example, we obtain the bottom left value by training on \texttt{Dist-Shift} and testing on \texttt{Inverse}.   }
    \label{fig:adaptivity-ablation}
\end{figure*}

\textbf{Adaptivity}. \Cref{fig:adaptivity-ablation} shows the evasion rate and text quality of our methods trained in the adaptive and non-adaptive settings when the provider uses \texttt{Llama2-13b} and the attacker uses \texttt{Llama2-7b}. 
We find that all adaptive attacks have an evasion rate of at least $96.6\%$, while the non-adaptive attacks have an evasion rate of at least $94.3\%$.
We achieve the highest overall evasion rate when training against the \texttt{Exp} watermark, which achieves an evasion rate of at least $97.0\%$. 
We train one attack, denoted \texttt{All}, against all four surveyed watermarking methods and test it against each watermark separately.
Interestingly, \texttt{All} performs slightly worse compared to training only on \texttt{Exp}, exhibiting an evasion rate of at least $96.3\%$ and a lower paraphrased text quality of at least $0.893$ (versus $0.901$ when training only on \texttt{Exp}).  
In summary, \Cref{fig:adaptivity-ablation} shows that adaptive attacks trained against one watermark remain highly effective when tested against unseen watermarks in the non-adaptive setting. 

\begin{figure*}
    \centering
    \includegraphics[width=0.7\linewidth]{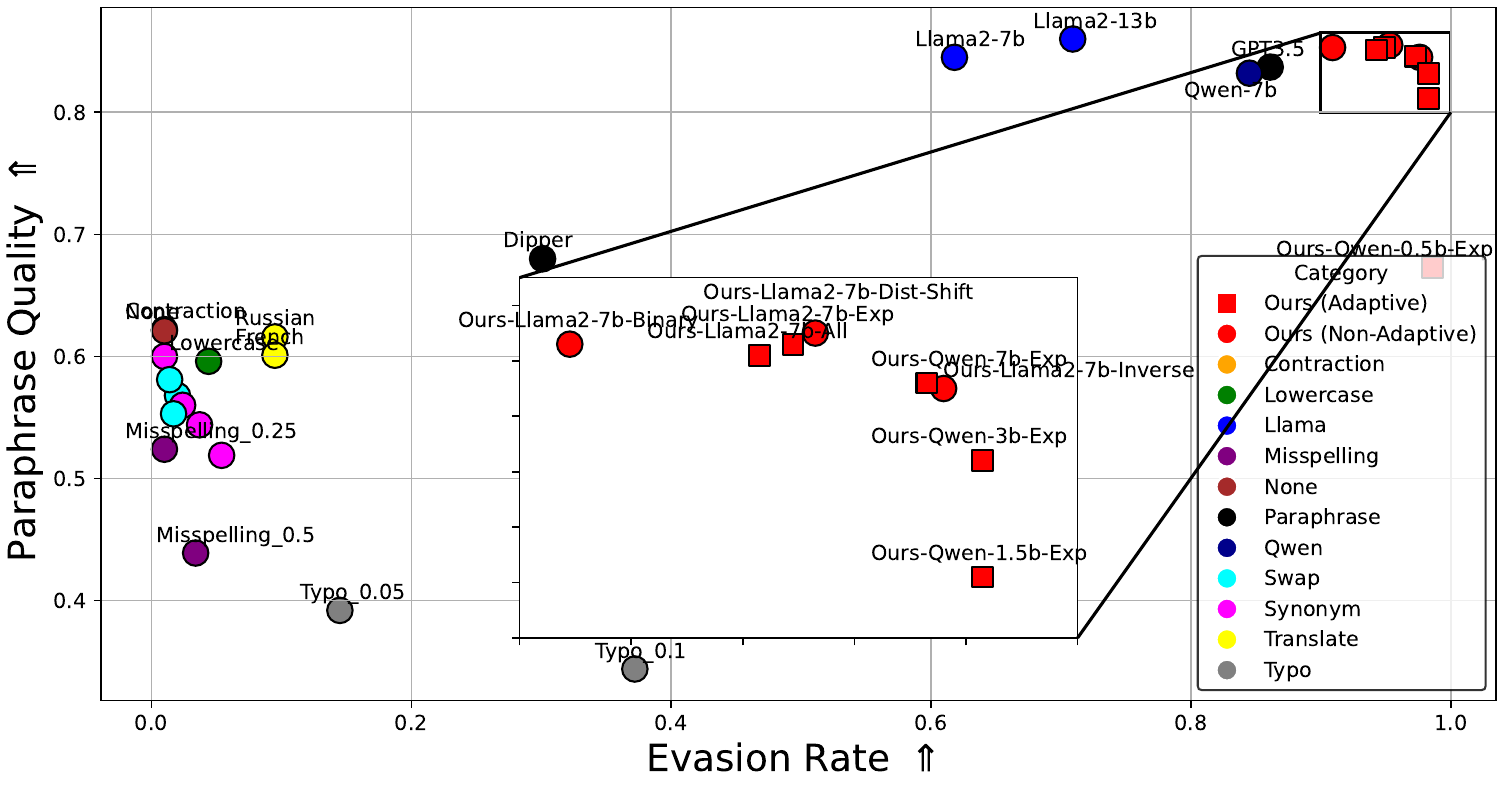}
    \caption{
    Adaptive attacks are Pareto-optimal. 
    We show the evasion rate versus text quality trade-off against the \texttt{Exp}~\citep{aaronson2023watermarking} watermark, corresponding to $(\epsilon,\delta)$-robustness from Eq. \ref{eq:robustness}.
    The provider uses a \texttt{Llama3.1-70b} model, whereas our attacker's models are up to $46\times$ smaller. 
    Non-adaptive attacks are marked by circles (\protect\whitecircle), adaptive attacks by squares (\protect\whitesquare). 
    Notation ``\texttt{Ours-Qwen-3b-Exp}'' means that we evaluate our attack using a \texttt{Qwen2.5-3b} model that was adaptively optimized against the \texttt{Exp} watermark. 
    }
    \label{fig:attack-summary-versus-dist-shift-70b}
\end{figure*}

\textbf{Model Sizes.} \Cref{fig:attack-summary-versus-dist-shift-70b} shows the Pareto front against the \texttt{Exp} watermark with a \texttt{Llama3.1-70b} target model.
Our attacker uses paraphraser models with at most 7b parameters, which is less than the 11b DIPPER model~\citep{krishna2023paraphrasing} currently used to test robustness.

We observe that
\begin{enumerate*}[label=(\arabic*), before={\unskip{ }}, itemjoin={{, }}, itemjoin*={{, and }}]
    \item \textbf{Non-adaptive baseline} attacks such as Contraction, Swapping and Synonym replacements are ineffective and have a low evasion rate of less than $20\%$.
    \item \textbf{Non-adaptive model-based} paraphrasing attacks such as using vanilla \texttt{Llama2-7b} or \texttt{ChatGPT3.5} models have a substantially higher evasion rate of $61.8\%$ up to $86.1\%$ respectively.
    Tuning \texttt{Llama2-7b} using our approach in the non-adaptive setting improves the evasion rate substantially to $90.9\%$ (when trained on \texttt{Binary}) and up to $97.6\%$ (when trained on \texttt{Inverse}).
    These non-adaptive, optimized attacks have a paraphrased text quality of 0.853 and 0.845, slightly improving over \texttt{ChatGPT3.5}, rated only 0.837. 
    \item \textbf{In the adaptive setting}, our fine-tuned \texttt{Qwen2.5-7b} achieves an evasion rate of $97.3\%$ at the highest text quality of $0.846$ compared to \texttt{Llama2-7b-Inverse}.
\end{enumerate*}

By ablating over \texttt{Qwen2.5} between 0.5b and 7b parameters, we find that attackers can strictly improve paraphrased text quality at similar evasion rates by using more capable paraphrases with more parameters. 
\Cref{fig:attack-summary-versus-dist-shift-13b} in the Appendix shows results against a \texttt{Llama2-13b} target model, which are consistent with those against \texttt{Llama3.1-70b}. 
Against smaller target models, attackers can achieve higher evasion rates and text quality ratings. 

\textbf{Text Quality}. \Cref{tab:qualitative_analysis} shows (i) a watermarked text sample generated using \texttt{Llama2-13b} with \texttt{Dist-Shift}, (ii) paraphrased text using a non-optimized \texttt{Llama2-7b} model, and (iii) paraphrased text obtained with an adaptively tuned \texttt{Llama2-7b} model using our attack.  
We observe that all paraphrased texts preserve quality, but our attack achieves the lowest green-to-red token ratio (i.e., maximizes the evasion rate). 
\Cref{tab:text-quality-metrics-appendix-full} in the Appendix shows a quantitative analysis of the median quality of generated text for a vanilla \texttt{Llama2-7b} model compared to our best adaptive and non-adaptive attacks.
It shows that text quality is preserved across five text quality metrics when using our attacks. 
We only show one paragraph of generated text that we truncated due to space restrictions and \Cref{tab:text_sample_full,tab:failed_sample_full} in the Appendix show non-truncated samples. 
\Cref{tab:failed_sample_full} shows a rare, cherrypicked example where our attack fails at evading watermark detection after paraphrasing. 

\begin{table*}[ht]
\centering
\begin{tabularx}{\textwidth}{X|X|X} 
\toprule
\small \texttt{Llama2-13b-}\texttt{Dist-Shift} &\small \texttt{Llama2-7b} &  \small \texttt{Ours-Llama2-7b-Dist-Shift}\\ 
\midrule
 \small
\rbox{C}%
\gbox{erv}%
\rbox{antes}%
\gbox{' "Don}%
\rbox{ Quixote}%
\gbox{" is a groundbre}%
\rbox{aking}%
\gbox{ masterpie}%
\rbox{ce}%
\gbox{ of literature that has enthr}%
\rbox{alled readers for}%
\gbox{ over 4}%
\rbox{0}%
\gbox{0}%
\rbox{ years.}%
\gbox{ The novel tells}%
\rbox{ the}%
\gbox{ story of Alonso Quixano, an old man who becomes so consumed by the chivalr}%
\rbox{ous}%
\gbox{ tales}%
\rbox{ of}%
\gbox{ his day that he ren}%
\rbox{ames himself}%
\gbox{ Don Qu}%
\rbox{ixote}%
\gbox{ and embarks upon an adventurous}%
\rbox{ journey}%
\gbox{ filled}%
\rbox{ with}%
\gbox{ humor, satire}%
\rbox{,}%
\gbox{ and tim}%
\rbox{eless}%
\gbox{ them}%
\rbox{es}%
\gbox{ about the}%
\rbox{ human}%
\gbox{ condition}%
\rbox{.}%
\gbox{$\hookleftarrow$$\hookleftarrow$ The story follows Quixote and}%
\rbox{ his}%
\gbox{ trustworthy}%
\rbox{ side}%
\gbox{kick}%
\rbox{ San}%
\gbox{cho}%
\rbox{ Panza, a simple and pragmatic s}%
\gbox{quire}%
\rbox{ who provides moments of humor and contrast to the lofty ideals of the delusional Qu}%
\rbox{ix}%
\gbox{ano/Don Quix}%
\rbox{ote}%
\gbox{. Their travels in rural Spain lead them to encounter all sorts of people they mistake for knights, ladies in dist}%
\rbox{ress}%
\gbox{, bandits, and giants who are in fact ordinary villagers going about their everyday lives.}
&
\small
\gbox{"Don}%
\rbox{ Quixote}%
\gbox{" is a groundbre}%
\rbox{aking}%
\gbox{ masterpie}%
\rbox{ce}%
\gbox{ of literature that has}%
\rbox{ capt}%
\gbox{ivated}%
\rbox{ readers for}%
\gbox{ over 4}%
\rbox{0}%
\gbox{0}%
\rbox{ years.}%
\gbox{ The novel tells}%
\rbox{ the}%
\gbox{ story of Alonso Quixano, an}%
\rbox{ elderly}%
\gbox{ man}%
\rbox{ whose fix}%
\gbox{ation on chivalr}%
\rbox{ous}%
\gbox{ tales leads}%
\rbox{ him to}%
\gbox{ change}%
\rbox{ his}%
\gbox{ name to Don}%
\rbox{ Quixote}%
\gbox{ and embark on a}%
\rbox{ thrilling}%
\gbox{ adventure replete}%
\rbox{ with}%
\gbox{ humor}%
\rbox{,}%
\gbox{ satire}%
\rbox{,}%
\gbox{ and tim}%
\rbox{eless}%
\gbox{ them}%
\rbox{es}%
\gbox{ concerning}%
\rbox{ the human condition.}%
\gbox{$\hookleftarrow$$\hookleftarrow$ The narr}%
\rbox{ative}%
\gbox{ follows Don Qu}%
\rbox{ixote}%
\gbox{ and}%
\rbox{ his}%
\gbox{ loyal squire}%
\rbox{ San}%
\gbox{cho}%
\rbox{ Pan}%
\gbox{se}%
\rbox{, a practical and}%
\gbox{ good}%
\rbox{-}%
\gbox{n}%
\rbox{ature}%
\gbox{d individual who}%
\rbox{ provides moments of}%
\gbox{ lev}%
\rbox{ity and contrast}%
\gbox{ to}%
\rbox{ the lofty ideals of the delusional Qu}%
\rbox{ix}%
\gbox{ano/Don Quix}%
\rbox{ote}%
\gbox{. Their travel}%
\rbox{s}%
\gbox{ across rural}%
\rbox{ Spain result}%
\gbox{ in encounters}%
\rbox{ with}%
\gbox{ various people}%
\rbox{ who}%
\gbox{ they mis}%
\rbox{ident}%
\gbox{ify as}%
\rbox{ knights, dist}%
\gbox{ressed}%
\rbox{ ladies,}%
\gbox{ bandits, and}%
\rbox{ ordinary}%
\gbox{ villagers going about their}%
\rbox{ daily lives.}%
&
 \small
\gbox{Spain}%
\rbox{ has been}%
\gbox{ bl}%
\rbox{essed with}%
\gbox{ a literary}%
\rbox{ tre}%
\gbox{asure}%
\rbox{ that has}%
\gbox{ captivated}%
\rbox{ readers for}%
\gbox{ over}%
\rbox{ four centuries in the}%
\gbox{ form of Miguel}%
\rbox{ de Cervantes}%
\gbox{'}%
\rbox{ immortal master}%
\gbox{pie}%
\rbox{ce}%
\gbox{, "Don}%
\rbox{ Quixote}%
\gbox{".}%
\rbox{ This}%
\gbox{ tim}%
\rbox{eless t}%
\gbox{ome we}%
\rbox{aves a hil}%
\gbox{arious and poignant tale}%
\rbox{ of a man consumed by}%
\gbox{ the chivalr}%
\rbox{ous}%
\gbox{ stories}%
\rbox{ of}%
\gbox{ his youth}%
\rbox{, who}%
\gbox{ ren}%
\rbox{ames}%
\gbox{ himself}%
\rbox{ Don Quix}%
\gbox{ote}%
\rbox{ and}%
\gbox{ sets}%
\rbox{ forth}%
\gbox{ on a journey filled}%
\rbox{ with moments}%
\gbox{ of}%
\rbox{ satire and a}%
\gbox{ piercing exam}%
\rbox{ination of}%
\gbox{ the}%
\rbox{ human}%
\gbox{ condition.}%
\rbox{$\hookleftarrow$$\hookleftarrow$}%
\gbox{As Don}%
\rbox{ Quix}%
\gbox{ote}%
\rbox{ and}%
\gbox{ his}%
\rbox{ trust}%
\gbox{y}%
\rbox{ s}%
\gbox{quire}%
\rbox{ Sancho}%
\gbox{ Panza traverse rural}%
\rbox{ Spain, they}%
\gbox{ encounter}%
\rbox{ various}%
\gbox{ uns}%
\rbox{uspect}%
\gbox{ing}%
\rbox{ villagers}%
\gbox{,}%
\rbox{ mistaking}%
\gbox{ them}%
\rbox{ for kn}%
\gbox{ights}%
\rbox{, maidens}%
\gbox{ in dist}%
\rbox{ress, bandits, and}%
\gbox{ even}%
\rbox{ giants.}
\\ \midrule
163 green and 36 red & 125 green and 69 red & 81 green and 78 red
\\ \bottomrule
\end{tabularx}
\caption{\label{tab:qualitative_analysis}(Left) Watermarked text from the provider's \texttt{Llama2-13b} model, (Center) a paraphrased version from a vanilla \texttt{Llama2-7b} model, and (Right) paraphrased text using our adaptively tuned \texttt{Llama2-7b} model.
Green/red indicates whether a token is watermarked.
A lower green-to-red token ratio implies a higher evasion rate. 
Due to space constraints, we only show truncated texts. 
\Cref{tab:text_sample_full,tab:failed_sample_full} in the Appendix show entire samples with up to 512 tokens.}
\end{table*}

\begin{figure*}
    \centering
    \begin{subfigure}{.49\linewidth} 
        \centering
        \includegraphics[width=1\linewidth]{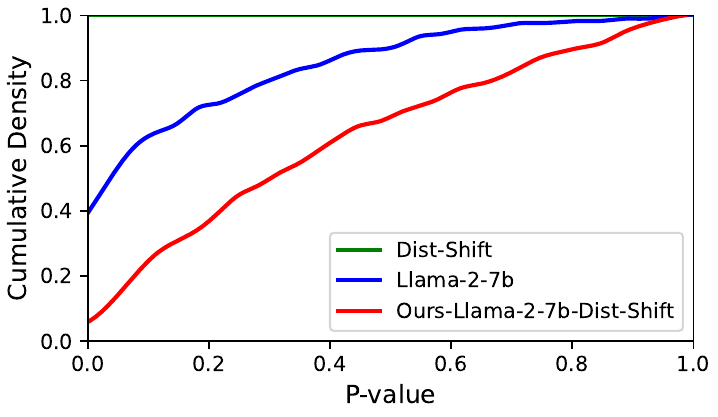}
    \end{subfigure}
    \hfill
    \begin{subfigure}{.49\linewidth} 
        \centering
        \includegraphics[width=1\linewidth]{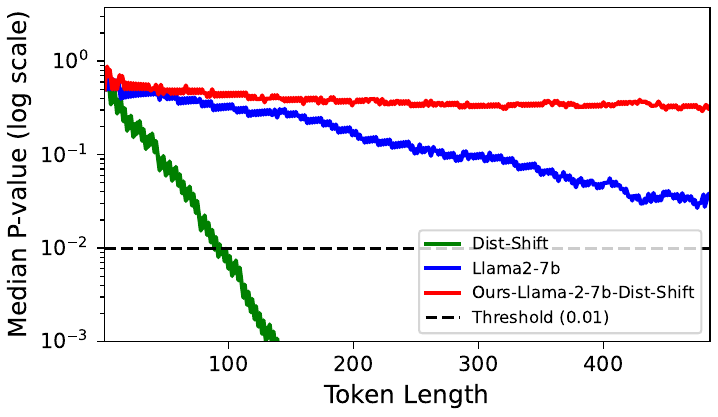}
    \end{subfigure}
    \hfill
    \caption{(Left) The cumulative density of p-values on the \texttt{Dist-Shift} watermark (green), a vanilla \texttt{Llama2-7b} paraphraser (blue) and our adaptive \texttt{Llama2-7b} paraphraser (red). (Right) The median p-value relative to the text token length with a threshold of $\rho=0.01$ (dashed line).  }
   \label{fig:detailed-analysis}
\end{figure*}

\textbf{Adaptive vs Non-adaptive.}
\Cref{fig:detailed-analysis} shows two results to compare the non-optimized \texttt{Llama2-7b} with our adaptively tuned \texttt{Llama2-7b} model.
The result on the left plots the cumulative density of p-values.
Our method strictly improves over the non-optimized model as it generates paraphrased text with higher mean p-values for watermark detection. 
The result on the right plots the expected p-value against the token length.
The watermarked text has a median p-value of less than $0.01$ after approximately 170 tokens, whereas the non-optimized \texttt{Llama2-7b} model has an expected p-value of 0.10 at around 500 tokens compared to an expected p-value of 0.4 for our adaptively tuned model. 

\textbf{Additional Testing.} {We present more results to compare adaptive versus non-adaptive attacks in \Cref{appendix:baseline_tests}, including tests against other recently released watermarking methods. 
These results are consistent with our findings in the main part of the paper that adaptive attacks are Pareto optimal and outperform much larger, closed-source systems such as GPT-4o at evading watermark detection.
We kindly refer the reader to \Cref{appendix:baseline_tests} for more baseline tests, \Cref{appendix:additional-statistics} for further statistical insights, and \Cref{sec:token_dist} for an analysis of the impact of paraphrasing on the top-50 token distribution. }

\section{Discussion}
\label{sec:discussion}

\textbf{Effectiveness of Adaptive Attacks.} Our work demonstrates that content watermarking methods for LLMs are vulnerable to adaptively optimized attacks. 
Attackers can adaptively fine-tune relatively small open-weight models, such as \texttt{Llama2-7b}~\citep{touvron2023llama}, in less than seven GPU hours to evade watermark detection from larger and more capable models, such as \texttt{Llama3.1-70b}~\citep{dubey2024llama}. 
Our attacks remain effective even in the non-adaptive setting when testing with unseen watermarking methods. 
Our findings challenge the robustness claims of existing watermarking methods, and we propose improved methods to test robustness using adaptive attacks. 

{\textbf{Analysis.} Studying \emph{why} adaptive attacks work is challenging due to the non-interpretability of the optimization process. 
The ability to maximize \Cref{eq:attack_objective} implies the ability to evade detection since \Cref{eq:attack_objective} encodes robustness for any watermarking method. 
The effectiveness of non-adaptive attacks could be explained by the fact that all surveyed watermarks are similar in that they operate on the token level. 
Hence, an effective attack against one watermark likely generalizes to other unseen watermarks.
Adaptive attacks further improve effectiveness as there are at least three learnable signals for paraphrasing watermarked text: (1) Avoid repeating token sequences, as they likely contain the watermark; (2) find text replacements with low impact on text quality to maximize the evasion rate (e.g., uncommon words or sentence structures); and (3) calibrate the minimum token edit distance and lexical diversity that, on average (over the randomness of the key generation process), evades detection.
We refer to \Cref{appendix:detailed_analysis} for a more detailed analysis of our approach's effectiveness.} 

\textbf{Attack Runtime.} Our attacks involve two steps: Dataset Curation and Model Optimization. 
Curating $7\,000$ samples requires less than 5 GPU hours, and model optimization requires only approximately 2 GPU hours for a \texttt{Llama2-7b} model at 16-bit precision. 
These attacks can be executed with limited computational resources and cost less than \$10 USD with current on-demand GPU pricing. 

\textbf{Restricted Attackers.} \citet{zhang2024watermarks} show that \emph{strong} watermarking, which resists any attack, is provably impossible under certain conditions. 
Our work instead focuses on robustness against restricted attackers with limited capabilities, such as limited compute resources, and we study whether robustness can be achieved in this setting. 
We show that current watermarks do not achieve robustness, and that even restricted attackers can evade detection at low costs. 

\textbf{Online Attacks.} Our work focuses on \emph{offline} attacks that do not require any access to the provider's watermark detection functionality. 
Offline attacks evaluate the robustness of a watermark without any information about the secret key generated by the provider. 
An \emph{online} attacker can learn information about the provider's secret key through accessing \texttt{Verify}, which reduces the attack's uncertainty and could substantially improve the attack's effectiveness further. 

\textbf{Limitations.}
Our study also did not focus on evaluating adaptive defences that could be designed against our adaptive attacks. 
Adaptive defences have not yet been explored, and we advocate studying their effectiveness.
{We believe our optimizable, adaptive attacks will enhance the robustness of future watermarking methods by including them in their design process, for instance, by using adversarial training.}
We focused exclusively on text generation tasks and did not explore other domains, such as source code generation or question-answering systems, where different text quality metrics may be used to evaluate an attack's success.
We did not consider the interplay between watermarking and other defenses, such as safety alignment or content filtering, which could collectively control misuse.

{We acknowledge that LLM-as-a-Judge is an imperfect and noisy metric that may not align with human judgment.  
In the main part of our paper, we use \texttt{Llama3-8B}-as-a-Judge, since this metric is easily reproducible. \Cref{appendix:baseline_tests} shows results using GPT-4o-mini-as-a-Judge, which are consistent with our findings.  
More work is needed to study the metric's alignment with human judgment. }

\section{Conclusion}
\label{sec:conclusion}
Our work demonstrates that current LLM watermarking methods are not robust against adaptively optimized attacks.
Even resource-constrained attackers can reliably ($\geq96.7\%$) evade detection with computational resource costs of $\leq$\$10 USD. 
They can achieve this with open-weight models that are $46\times$ smaller than the provider's model. 
Even in the non-adaptive settings, our adaptively tuned attacks outperform all other surveyed attacks, including paraphrasing with substantially larger models such as OpenAI's GPT4o.
Our findings challenge the security claims of existing watermarking methods and show that they do not hold even against resource-constrained attackers.
We suggest that future defenses must consider adaptive attackers to test robustness.

\section*{Impact Statement}
This work investigates the robustness of watermarking methods for large language models (LLMs), which has implications for content authentication and the responsible deployment of AI systems. 
Our findings demonstrate that attackers with limited computational resources can undermine the robustness of current watermarking methods by using adaptively optimized attacks.
This vulnerability could have societal implications as major AI providers increasingly adopt watermarking to promote responsible AI use and control misuse, including the proliferation of LLM-generated misinformation and online spam.

By publicly releasing our methods, findings, and source code, we hope to encourage the development of more robust watermarking methods that can better withstand adaptive attacks (e.g., by increasing the computational complexity of such attacks or making them less effective). 
We acknowledge that our research could potentially be misused to evade existing deployments of watermarks.  
However, these deployments are still in experimental phases. 
We believe the benefit of releasing our work outweighs the potential harm and hope that our work inspires the development of more robust content authentication methods.

\bibliography{references}
\bibliographystyle{icml2025}


\appendix
\onecolumn
\section{Appendix}

\subsection{Quality Metrics}
\label{appendix:metrics}

Ideally, to evaluate the quality of an LLM-generated text, one would need a set of human evaluators, each giving their own score according to a certain rubric, and then have all the scores aggregated. However, this is impractical for both the attacker and the defender. Therefore, we employ multiple surrogate metrics from the literature: LLM-Judge, LLM-CoT, and LLM-Compare from \citep{piet2023mark}, Mauve~\citep{pillutla2021mauvemeasuringgapneural}, and Perplexity (PPL) with \texttt{Llama3-8B-Instruct}. Note that all of these are implemented in the \texttt{MarkMyWords (MMW)} \citep{piet2023mark} benchmark utilized for our experiments. All the metrics evaluate a response (whether watermarked or a perturbed sample) against a baseline (either the original prompt, a non-watermarked sample, or the model's logit distribution). Below is a description of each metric, along with an indication of whether higher or lower values are better for that metric.

\paragraph{LLM-Judge:} LLM-Judge directly prompts an instruction-tuned large language model to evaluate the quality of a certain response with respect to its original prompt. The response is decoded greedily (temperature =0) to ensure deterministic results. Criteria evaluated include: accuracy, level of detail, and typographical, grammatical, and lexical correctness. A higher score is better. For this, we use \texttt{Llama3-8B-Instruct} to evaluate, using the following prompt:


\begin{tcolorbox}[
    colframe=chatgptcolor, 
    title=\textbf{LLM-Judge Prompt}, 
    sharp corners,
    boxrule=1pt, 
    width=\linewidth, 
    fonttitle=\bfseries, 
    coltitle=black, 
]
\small\texttt{You are given a prompt and a response, and you provide a grade out of 100 measuring the quality of the response, in terms of accuracy, level of details, and typographical, grammatical and lexical correctness. 
Remove points as soon as one of the criteria is missed.
Prompt:\{\}, Response:\{\}}
\end{tcolorbox}

\paragraph{LLM-CoT (Chain-of-Thought):} LLM-CoT evaluates the quality of the watermarked/attacked responses using CoT-based reasoning~\cite{wei2023chainofthoughtpromptingelicitsreasoning}. A higher score is better. For this, we also use \texttt{Llama3-8B-Instruct} to evaluate, using the following prompt:

\begin{tcolorbox}[
    colframe=chatgptcolor, 
    title=\textbf{LLM-CoT Prompt}, 
    sharp corners,
    boxrule=1pt, 
    width=\linewidth, 
    fonttitle=\bfseries, 
    coltitle=black, 
]
\small\texttt{Please act as an impartial judge and evaluate the quality of the response provided by an AI assistant to the user question displayed below. Your evaluation hinges on the following factors: accuracy, level of details, typographical correctness, grammatical correctness and lexical correctness. Be as objective as possible, and account for every one of these criteria in our evaluation. Begin your evaluation by providing a short explanation.  After providing your explanation, please rate the response on a scale of 1 to 100 by strictly following this format: "[[rating]]", for example: "Rating: [[51]]". Your rating must come AFTER your explanation, not before. Prompt:\{\}, Response:\{\} }
\end{tcolorbox}

\paragraph{LLM-Comparator:} The LLM-Comparator is used to compare non-watermarked baseline response and the watermarked or attacked response. 0 indicates that the non-watermarked response is better, 0.5 indicates a tie, and 1 shows that the watermarked or attacked response is better. For this, we also use \texttt{Llama3-8B-Instruct} to evaluate, using the following prompt:

\begin{tcolorbox}[
    colframe=chatgptcolor, 
    title=\textbf{LLM-Comparator Prompt}, 
    sharp corners,
    boxrule=1pt, 
    width=\linewidth, 
    fonttitle=\bfseries, 
    coltitle=black, 
]
\small\texttt{ Please act as an impartial judge and evaluate the quality of the responses provided by two AI assistants to the user question displayed below. You should choose the assistant that follows the user's instructions and answers the user's question better. Your evaluation should consider factors such as the helpfulness, relevance, accuracy, depth, creativity, and level of detail of their responses. Also account for typographical correctness, grammatical correctness and lexical correctness. Begin your evaluation by comparing the two responses and provide a short explanation. Avoid any position biases and ensure that the order in which the responses were presented does not influence your decision. Do not allow the length of the responses to influence your evaluation. Do not favor certain names of the assistants. Be as objective as possible. After providing your explanation, you must output your final verdict by strictly following this format: * "[[A]]" if assistant A is better, * "[[B]]" if assistant B is better, and * "[[C]]" for a tie. For example, "Verdict: [[C]]". Prompt: \{\}, [[Start of Assistant A]] \{\} [[End of Assistant A's Answer]], [[Start of Assistant B]] \{\} [[End of Assistant B's Answer]] }
\end{tcolorbox}

\paragraph{MAUVE:} MAUVE measures the similarity between two text distributions. 
In our case, the two distributions are the non-watermarked baseline response and the watermarked/paraphrased response.
Higher MAUVE scores indicate that both texts match their content, quality and diversity. 
MAUVE is computed with the Kullback-Leibler (KL) divergences between two distributions in a lower-dimensional latent space. 
It correlates with human evaluations over baseline metrics for open-ended text generation~\cite{pillutla2021mauvemeasuringgapneural}. We use the \texttt{gpt2-large} model to compute the MAUVE score in our implementation.

\paragraph{Perplexity (PPL):} Perplexity is a common language modelling metric that quantifies how well a model predicts a text sample. 
It is calculated based on the probability that the model assigns to a sequence of words. Lower perplexity values indicate that the model is more confident and accurate in its predictions, making lower scores better for this metric.

\Cref{tab:text-quality-metrics-appendix-full} shows the median text quality metrics to compare the vanilla \texttt{Llama2-7b} paraphraser to our best adaptive and non-adaptive attacks against the \texttt{Llama2-13B} and \texttt{Llama3.1-70B} target models. The table shows that our attacks have similar quality to the vanilla \texttt{Llama2-7b} paraphraser across the board. Our attacks have a higher MAUVE score, indicating that our text is closer to the non-watermarked text than the vanilla \texttt{Llama2-7b} paraphraser. The higher perplexity is not a concern, as it merely indicates that the large language model does not expect the text.

\begin{table}[ht]
\centering
\begin{tabular}{@{}lccccc@{}}
\toprule
Target: \texttt{Llama2-13B}     & LLM-Judge $\Uparrow$ & LLM-CoT$\Uparrow$ & LLM-Compare$\Uparrow$ & Mauve$\Uparrow$ & PPL$\Downarrow$    \\ \midrule
\texttt{Llama2-7b}     & 0.92     & 0.85   & 0.00        & 0.17 & 4.74  \\
Ours-Best-Adaptive     & 0.92     & 0.85   & 1.00       & 0.42 & 6.69  \\
Ours-Best-Non-Adaptive & 0.92     & 0.85   & 0.50       & 0.37 & 6.32  \\
\midrule
 Target: \texttt{Llama3.1-70B}     &           &         &             &       &        \\ \midrule
\texttt{Llama2-7b}     & 0.95       & 0.72   & 0.00        & 0.22 & 4.84 \\
Ours-Best-Adaptive     & 0.95     & 0.72   & 0.50       & 0.55 & 6.10  \\
Ours-Best-Non-Adaptive & 0.95     & 0.72   & 0.50       & 0.31 & 6.15  \\ \bottomrule
\end{tabular}
\caption{
\label{tab:text-quality-metrics-appendix-full}Various median text quality metrics to compare the vanilla \texttt{Llama2-7b} paraphraser to our best adaptive and non-adaptive attacks. We limit all attacks to at most 7b parameter models.}
\end{table}

\subsection{Prompt-set Curation}\label{appendix:prompt-set-curation}

The \textbf{evaluation set} consists of 296 prompts from \citet{piet2023mark}, covering book reports, storytelling, and fake news. 
The \textbf{training set} comprises a synthetic dataset of 1\,000 prompts, covering diverse topics including reviews, historical summaries, biographies, environmental issues, science, mathematics, news, recipes, travel, social media, arts, social sciences, music, engineering, coding, sports, politics and health. 
To create this dataset, we repeatedly prompt a large language model (ChatGPT-4) to generate various topic titles.
These titles were then systematically combined and formatted into prompts.

The synthetic training dataset is non-overlapping with the evaluation set. 
Nonetheless, in realistic scenarios, it is plausible that an attacker might train and evaluate their paraphraser using the same dataset. 
Given the low cost of our attack (USD $\leq10 \$ $), attackers can easily train their own paraphrasers.

\subsection{Preference-data Curation}
\label{appendix:additional-results-dataset-curation}
\begin{wrapfigure}{R}{0.5\linewidth}
    \centering
    \includegraphics[width=\linewidth]{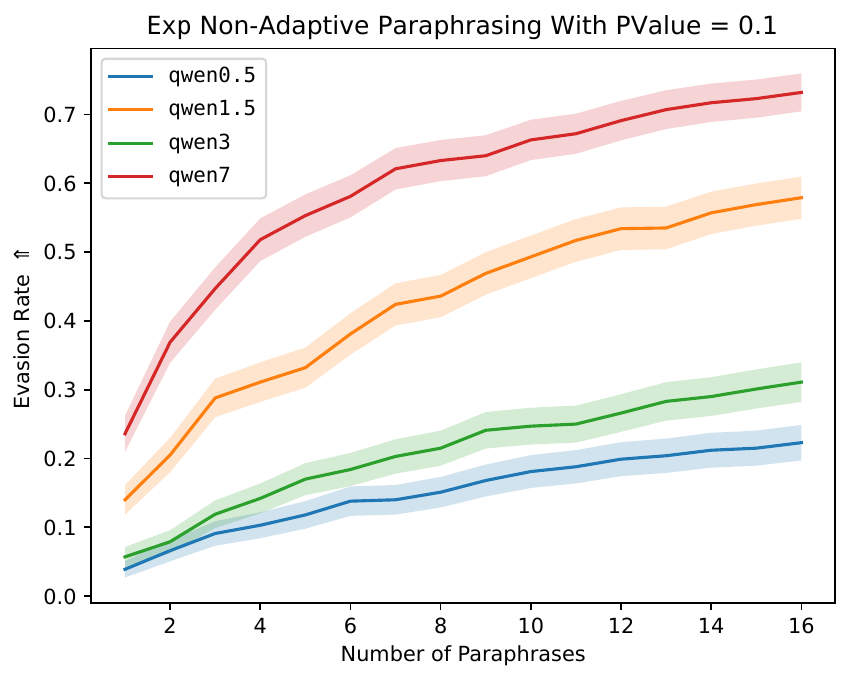}
    \caption{The expected evasion rate versus the repetition rate ablated over varying model sizes of \texttt{Qwen2.5}~\citep{qwen2.5} against the \texttt{Exp} watermark (\Cref{alg:preference-dataset-collection}, lines 9-12). Shaded areas denote 95\% confidence intervals.}     \label{fig:paraphrase-qwen0.05}
\end{wrapfigure}
For every prompt in the training set, we generate watermarked output using each watermark; then, we use that output as input to our paraphrasers. Each paraphraser generates 16 paraphrases for each input. We then filter these paraphrases as per \Cref{alg:preference-dataset-collection} to create the training preference pairs. Larger models have higher quality output and so have a higher yield of successful paraphrases. We use the same number of paraphrases for each model, even though they may generate different yields.

\Cref{fig:paraphrase-qwen0.05} shows the expected evasion rate versus the number of paraphrases ablated over varying model sizes of \texttt{Qwen2.5}~\citep{qwen2.5} against the \texttt{Exp} watermark. We find that the expected evasion rate increases with the number of paraphrases, but the rate of increase diminishes as the number of paraphrases increases. We find that the expected evasion rate does not improve significantly close to 16 paraphrases and that bigger models tend to have higher evasion rates for the same number of paraphrases. An exception to this is the 1.5b model, which surprisingly performs very well (better than the 3b) for the same number of paraphrases. This, however, could be due to different pretraining parameters of the base model or other factors.

\subsection{Baseline Testing against other Watermarks}
\label{appendix:baseline_tests}
{We include more robustness tests against recently released watermarks such as SynthID~\citep{dathathri2024scalable}, Unigram~\citep{zhao2024provable} and SIR~\citep{liu2024a}. We refer to the author's papers for detailed descriptions of these watermarks.
For this evaluation, we implemented our attack in the MarkLLM framework~\citep{pan-etal-2024-markllm}, used our \texttt{Qwen2.5-3b} paraphraser trained against the \textsc{EXP} watermark from the main part of the paper, and adaptively tuned a new \texttt{Qwen2.5-3b} paraphraser against the Unigram watermark.
For the purpose of quick evaluation, we limit the token length to 256 tokens, noting that, as shown in \Cref{fig:detailed-analysis}, the results are similar for longer texts.
GPT-4o is part of the Pareto front only against SIR and KGW due to its high text quality and low evasion rates of less than 90\%. 
It is not part of the Pareto front against SynthID, EXP and Unigram, where only our attacks are part of the Pareto front.
While it may be possible to use better prompts for GPT-4o to achieve a higher text quality or evasion rate, there are other limitations when using closed systems to evade detection.
\begin{enumerate}
    \item Their usage can be expensive as the user is typically charged per token.
    \item The system could embed its own watermark into paraphrased text.  
    \item There could be guardrails such as safety alignments which prevent these systems from arbitrarily paraphrasing text. 
\end{enumerate}
In contrast, our method allows working with relatively small open-weight models that adversaries can fully control.
}

\begin{figure}
    \centering
    \includegraphics[width=1.0\linewidth]{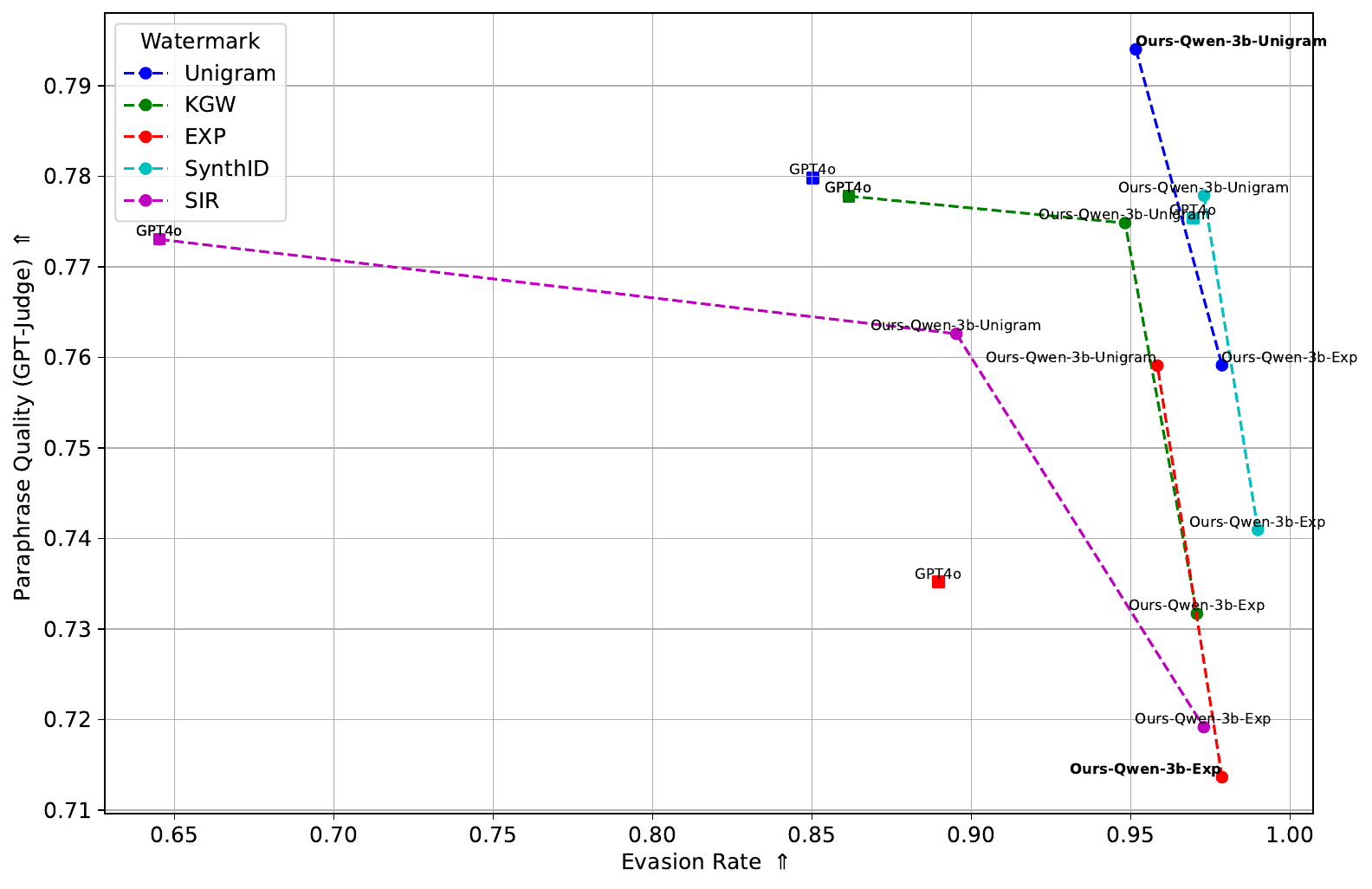}
    \caption{{Additional results using \texttt{Qwen2.5-3b} against KGW and EXP, which we study in the main part of the paper, and more recently released watermarks such as SynthID~\citep{dathathri2024scalable}, Unigram~\citep{zhao2024provable} and SIR~\citep{liu2024a}. Dashed lines denote the Pareto front, and we highlighted adaptively trained attacks in bold. 
    We used GPT-4o's version from November 23rd, 2024. 
    The y-axis uses GPT-4o-mini as a judge, and the x-axis shows the evasion rate.} 
    }
    \label{fig:further-testing-robustness}
\end{figure}

\begin{figure}
    \centering
    \includegraphics[
        width=1.0\linewidth]{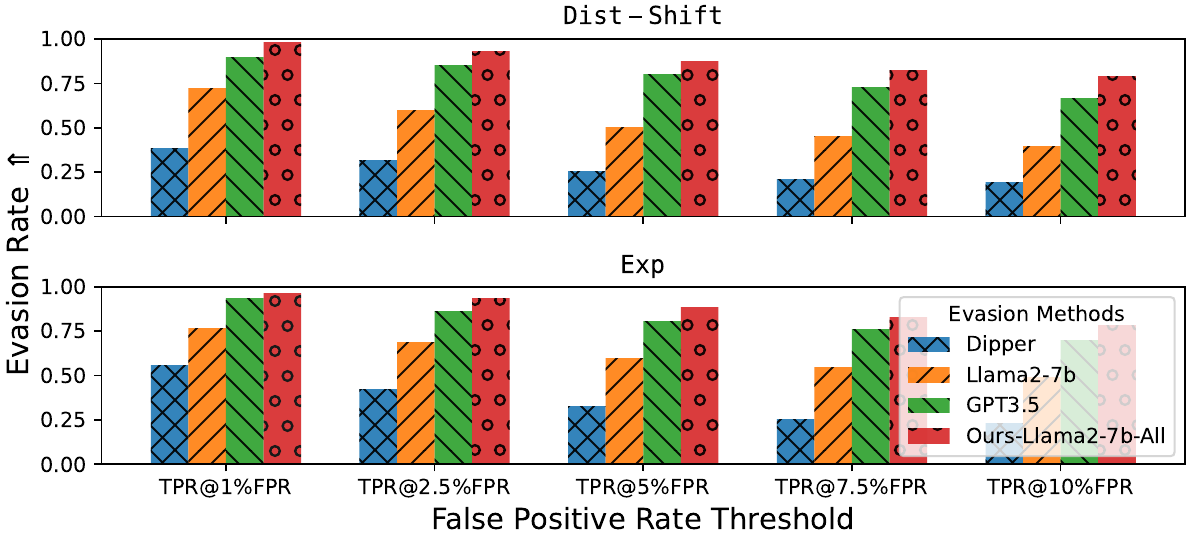}
    \caption{The evasion rates against a watermarked \texttt{Llama2-13b} model. We compare non-adaptive attacks, including ChatGPT3.5, versus our adaptively fine-tuned \texttt{Llama2-7b} paraphraser model. }
    \label{fig:false-positive-rates}
\end{figure}

 \subsection{Additional Statistics}
\label{appendix:additional-statistics}

We provide additional statistical insights complementing the robustness tests described in \Cref{appendix:baseline_tests}. For brevity and clarity, we illustrate the statistics primarily with the Unigram watermark~\citep{zhao2024provable}, noting similar results across other watermark methods.

\begin{figure}[ht]
    \centering
    \includegraphics[width=1.0\linewidth]{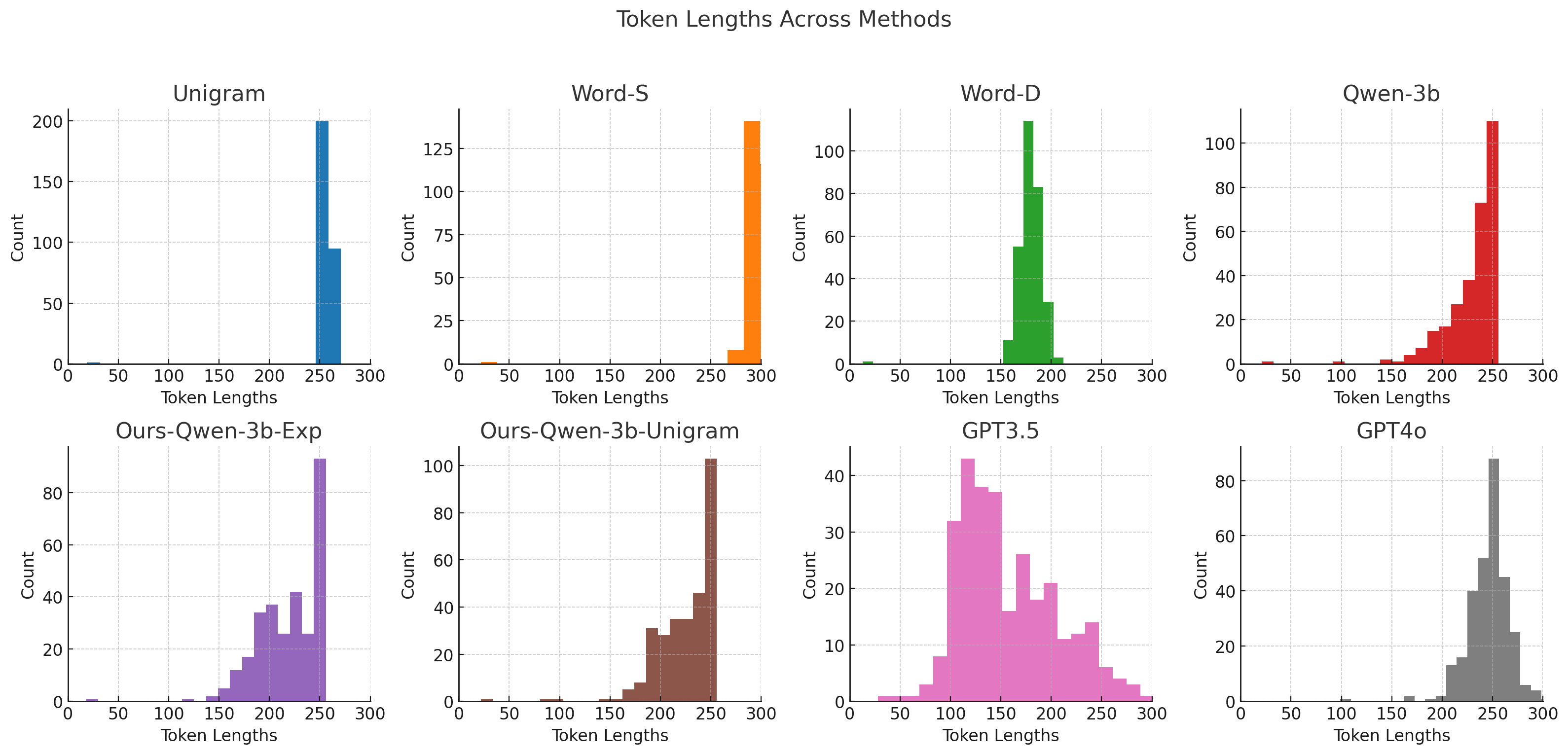}
    \caption{The distribution of the number of tokens in the watermarked text and the paraphrased texts. The x-axis shows the number of tokens, and the y-axis shows the number of samples.}
    \label{fig:token-lengths}
\end{figure}

\paragraph{Token Length Analysis}
\Cref{fig:token-lengths} shows the distribution of token lengths for watermarked texts and different perturbations. Our tuned paraphrasers (\texttt{Qwen2.5-3b-Unigram} and \texttt{Qwen2.5-3b-EXP}) produce slightly shorter paraphrases compared to the base \texttt{Qwen2.5-3b} model and the watermarked responses themselves. This reduction in length likely arises from the optimization objective, which does not explicitly penalize brevity. Such behavior could be adjusted by modifying the objective function when selecting positive and negative samples. Non-optimized methods exhibit varied token lengths: GPT-3.5 generates even shorter responses, GPT-4o produces relatively longer texts, while word substitution (Word-S) and deletion (Word-D) methods behave as expected, respectively increasing or decreasing token counts.

\begin{figure}[ht]
    \centering
    \includegraphics[width=1.0\linewidth]{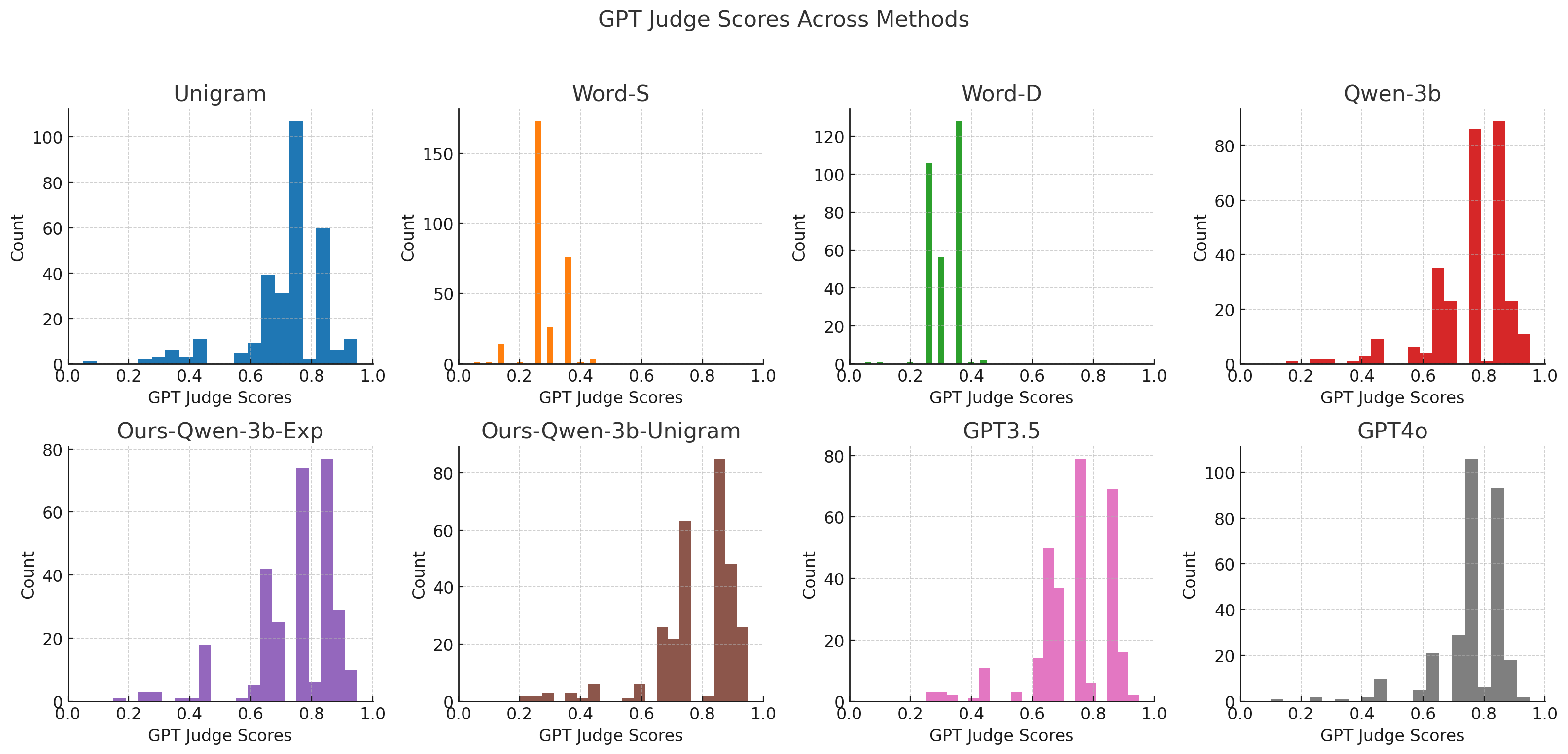}
    \caption{The distribution of the GPT-Judge scores for the watermarked text and the paraphrased texts. The x-axis shows the GPT-Judge score, and the y-axis shows the number of samples.}
    \label{fig:gpt-judge}
\end{figure}

\paragraph{GPT-Judge Quality Scores}
The GPT-Judge scores are evaluated using GPT-4o-mini. \Cref{fig:gpt-judge} indicates text quality across methods. Optimized paraphrasers (\texttt{Qwen2.5-3b-Unigram} and \texttt{Qwen2.5-3b-EXP}) exhibit similar high-quality scores to those of the unattacked Unigram watermark, base \texttt{Qwen2.5-3b}, GPT-3.5, and GPT-4o methods. In contrast, simple perturbations like Word-S and Word-D achieve significantly lower quality scores.

\begin{figure}[ht]
    \centering
    \includegraphics[width=1.0\linewidth]{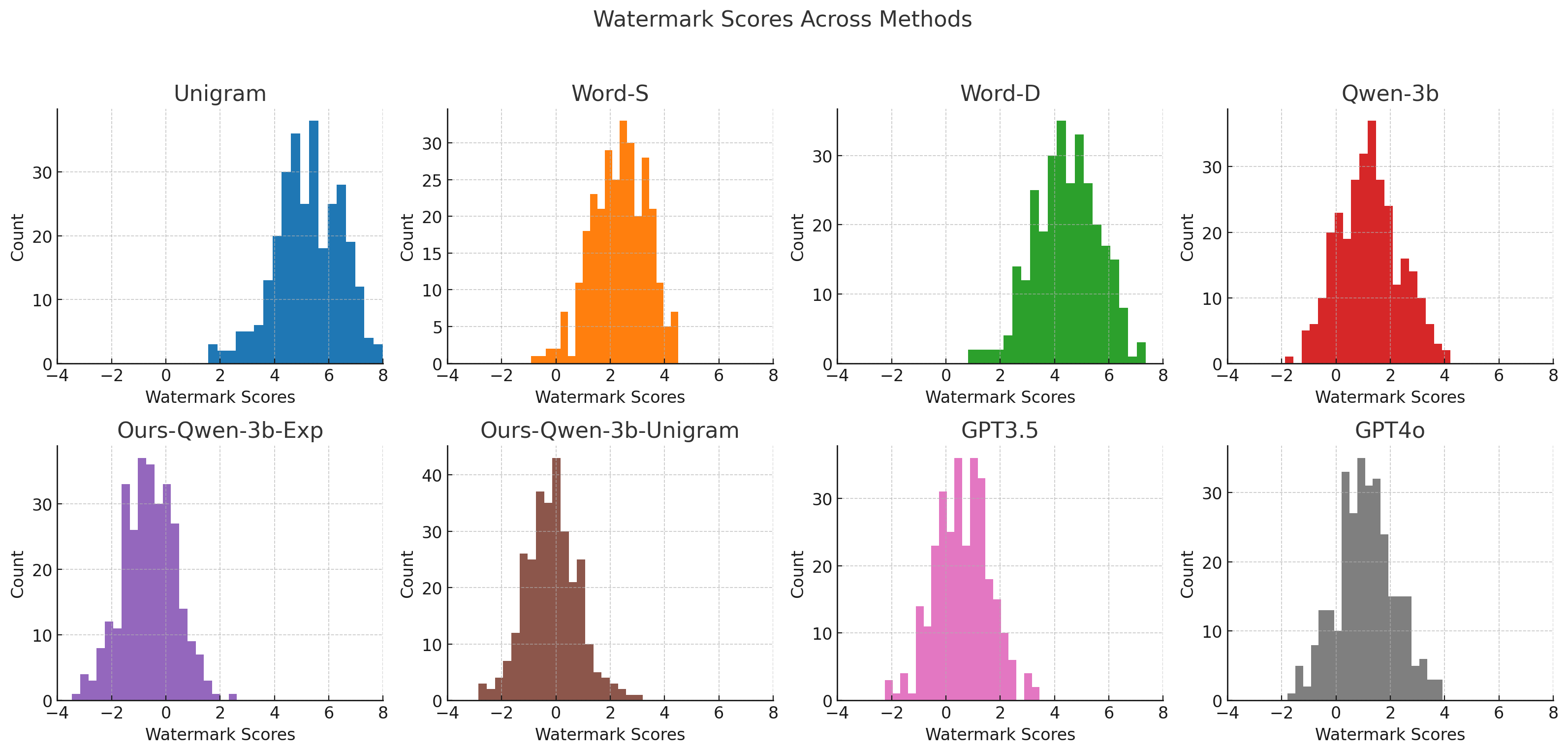}
    \caption{The distribution of the watermark scores for the watermarked text and the paraphrased texts. The x-axis shows the watermark score, and the y-axis shows the number of samples.}
    \label{fig:watermark-scores}
\end{figure}

\paragraph{Watermark Scores}
\Cref{fig:watermark-scores} illustrates watermark detection scores as measured by the MarkLLM framework~\citep{pan-etal-2024-markllm}. The unattacked watermarked texts have notably high scores (centered around 5). Simple perturbations, such as word deletions, have no impact, while word substitutions moderately reduce scores. Non-tuned paraphrasing methods (\texttt{Qwen2.5-3b}, GPT-3.5, GPT-4o) substantially lower watermark scores (centered around 1). Adaptively fine-tuned paraphrasers (\texttt{Qwen2.5-3b-EXP} and \texttt{Qwen2.5-3b-Unigram}) achieve the lowest scores, typically centered around -1, highlighting their effectiveness in evading detection.

\subsection{Token Distribution}
\label{sec:token_dist}

\begin{figure}[H]
    \centering
    \includegraphics[width=1.0\linewidth]{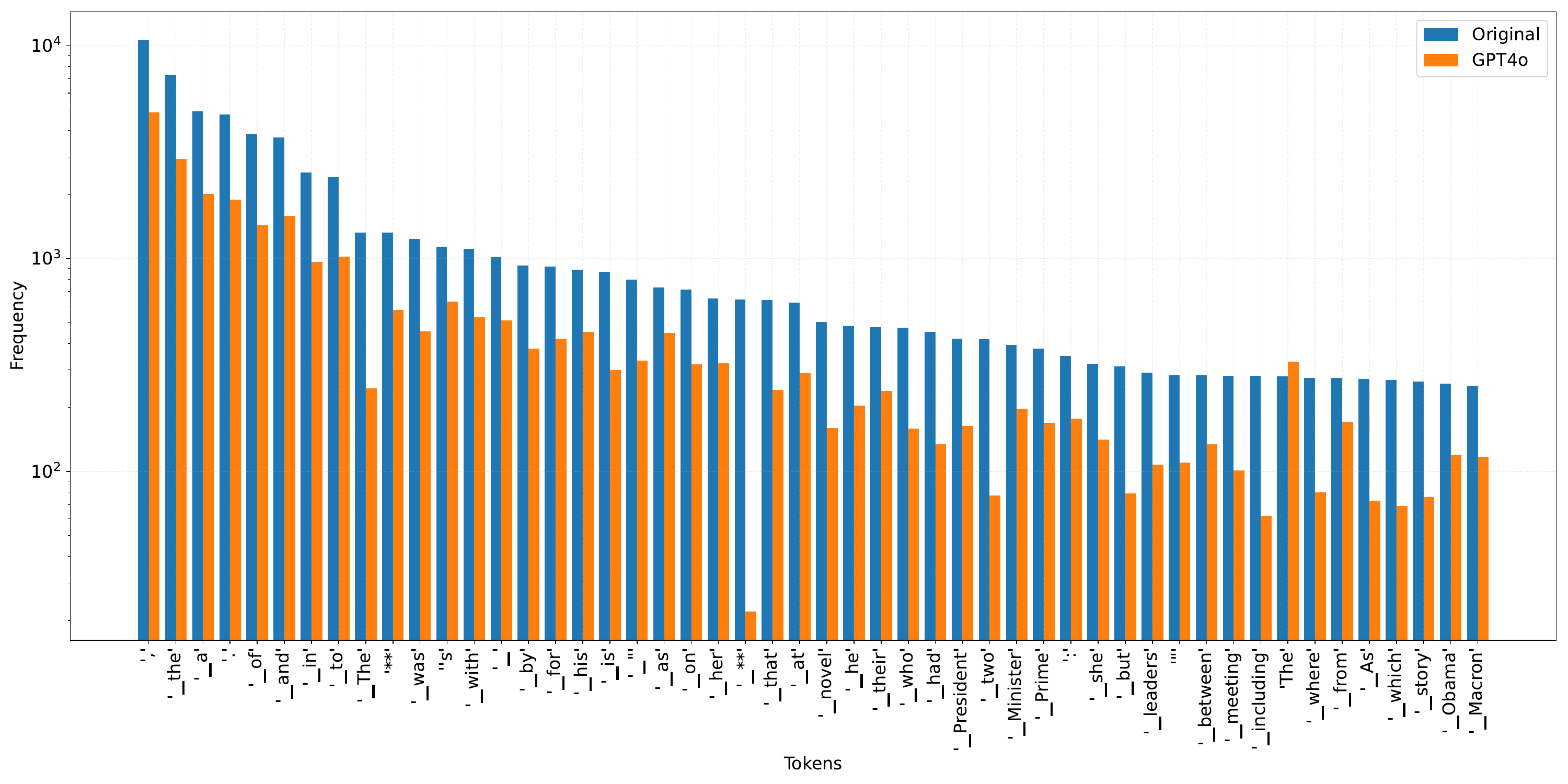}
    \caption{{An analysis of the top-50 tokens in paraphrased text generated with the Unigram watermark~\citep{zhao2024provable}, using GPT-4o as a paraphraser.}}
    \label{fig:detailed-token-analysis-gpt4o}
\end{figure}

\begin{figure}[H]
    \centering
    \includegraphics[width=1.0\linewidth]{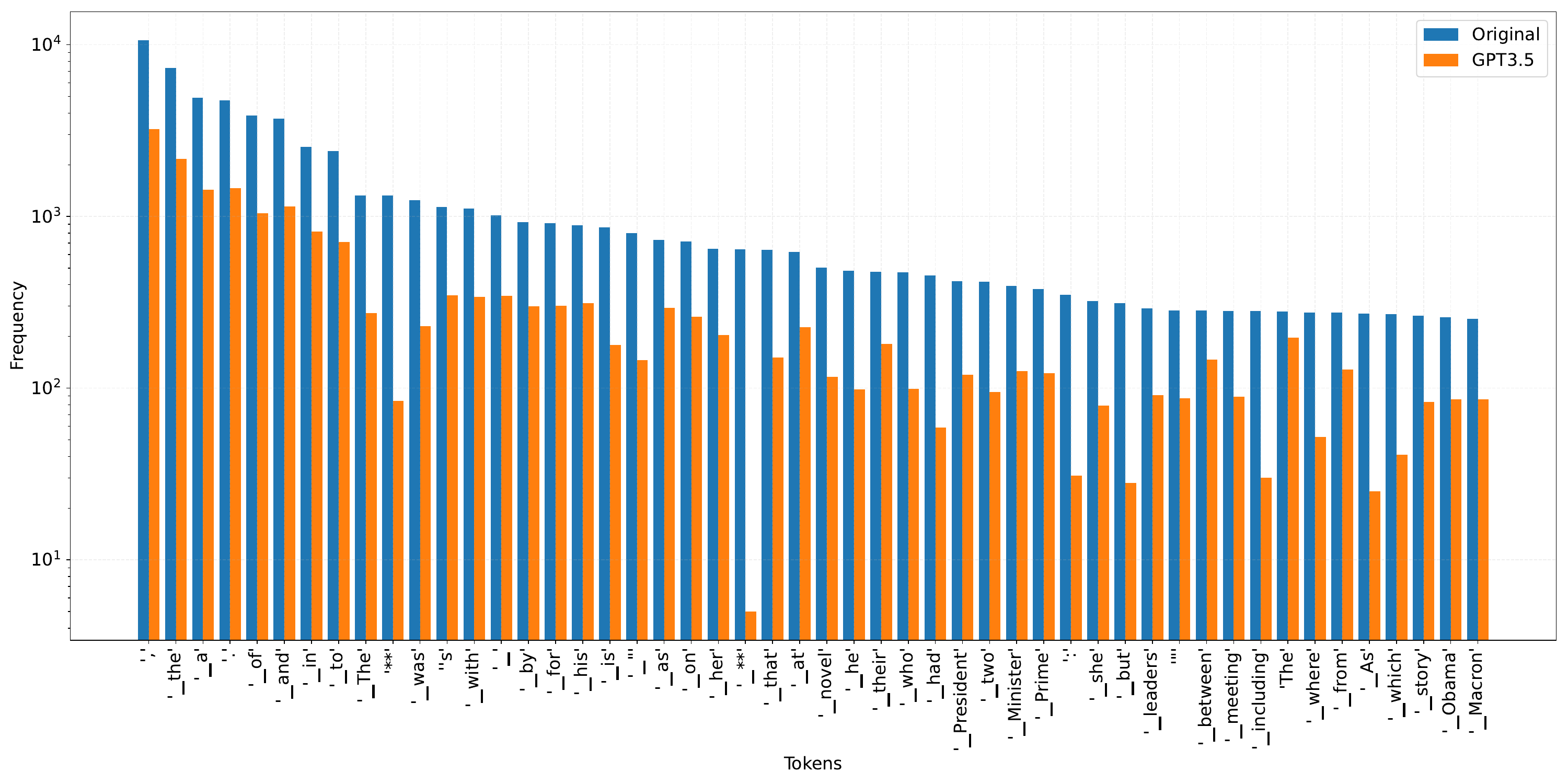}
    \caption{{An analysis of the top-50 tokens in paraphrased text generated with the Unigram watermark~\citep{zhao2024provable}, using an off-the-shelf \texttt{Qwen2.5-3b} model as a paraphraser.}}
    \label{fig:detailed-token-analysis-qwen}
\end{figure}

\begin{figure}[H]
    \centering
    \includegraphics[width=1.0\linewidth]{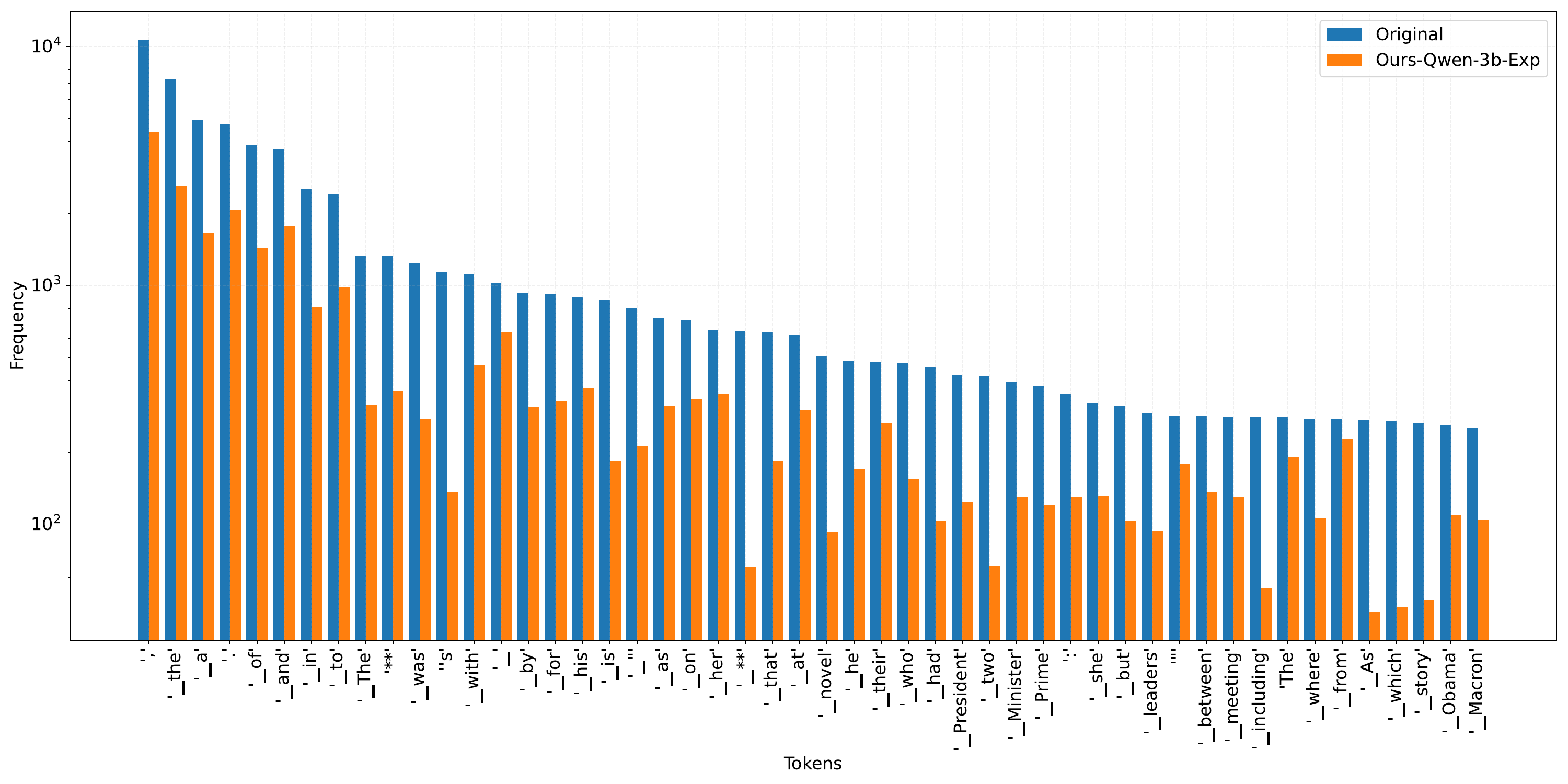}
    \caption{{An analysis of the top-50 tokens in paraphrased text generated with the Unigram watermark~\citep{zhao2024provable}, using our adaptively tuned \texttt{Qwen2.5-3b} model as a paraphraser.}}
    \label{fig:detailed-token-analysis-ours}
\end{figure}

{\textbf{Text Quality.} \Cref{fig:detailed-token-analysis-gpt4o,fig:detailed-token-analysis-qwen,fig:detailed-token-analysis-ours} show the top-50 token distribution that appear in the watermarked text. 
We compare it with the token frequency in the paraphrased text using as paraphrasers (i) GPT-4o, (ii) a baseline \texttt{Qwen2.5-3b} model and (iii) our adaptively tuned \texttt{Qwen2.5-3b} model against the Unigram watermark~\citep{zhao2024provable}. 
We observe that all paraphrasers have a similar token distribution and that across all three paraphrasers, on average, the top 50 tokens appear less frequently than in the original, watermarked text.
The largest difference we observe between the baseline \texttt{Qwen2.5-3b} and our adaptively tuned model are the frequencies of the tokens 'The' and ' ' (space between words), which our model uses less frequently. 
Compared to GPT-4o, the baseline \texttt{Qwen2.5-3b} model uses some tokens, such as ' As', less frequently, while other tokens, such as ' but', appear more frequently. 
}

\subsection{Detailed Textual Analysis}
\label{appendix:detailed_analysis}

{Our goal is to further analyze why our adaptively tuned paraphraser better evades detection than other approaches.
We begin by studying the overlap of N-grams between the watermarked and paraphrased texts, which we call the N-gram overlap ratio between two sequences $x_1,x_2 \in \mathcal{V}^*$.
}
\begin{align}
N_g(x_1,x_2,n) = \frac{\lvert \text{ngrams}(x_1, n) \cap \text{ngrams}(x_2,n) \rvert}{\lvert \text{ngrams}(x_1, n) \cup \text{ngrams}(x_2, n) \rvert}
\label{eq:n_gram_overlap}
\end{align}
{The 'ngrams' function tokenizes a sequence and returns the set of n-grams. 
The N-gram overlap ratio is always between [0,1]. 
A high overlap for a given $n\in \mathbb{N}$ indicates that the same N-grams appear in both sequences. 
Since the surveyed watermarks operate on a token level, a low overlap ratio would suggest a high evasion rate. We also evaluate the token edit distance ratio between two sequences, which is calculated as follows: 
}
\begin{align}
    L(x_1,x_2) =  \frac{\text{Levenshtein}(x_1, x_2)}{\text{len}(x_1) + \text{len}(x_2)}
\end{align}

{The token edit distance calculates the Levensthein distance between two sequences. Note that the N-gram overlap ratio is calculated over sets of N-grams. In contrast, the Levenshtein distance is calculated over (ordered) sequences, meaning that the position of the token matters. 
A high Token Edit Distance ratio suggests that two texts do not have the same tokens at the same positions in the sequence, which also suggests a higher evasion rate. }

\begin{figure}[H] 
    \centering
    \begin{subfigure}{.49\linewidth} 
        \centering
        \includegraphics[width=1\linewidth]{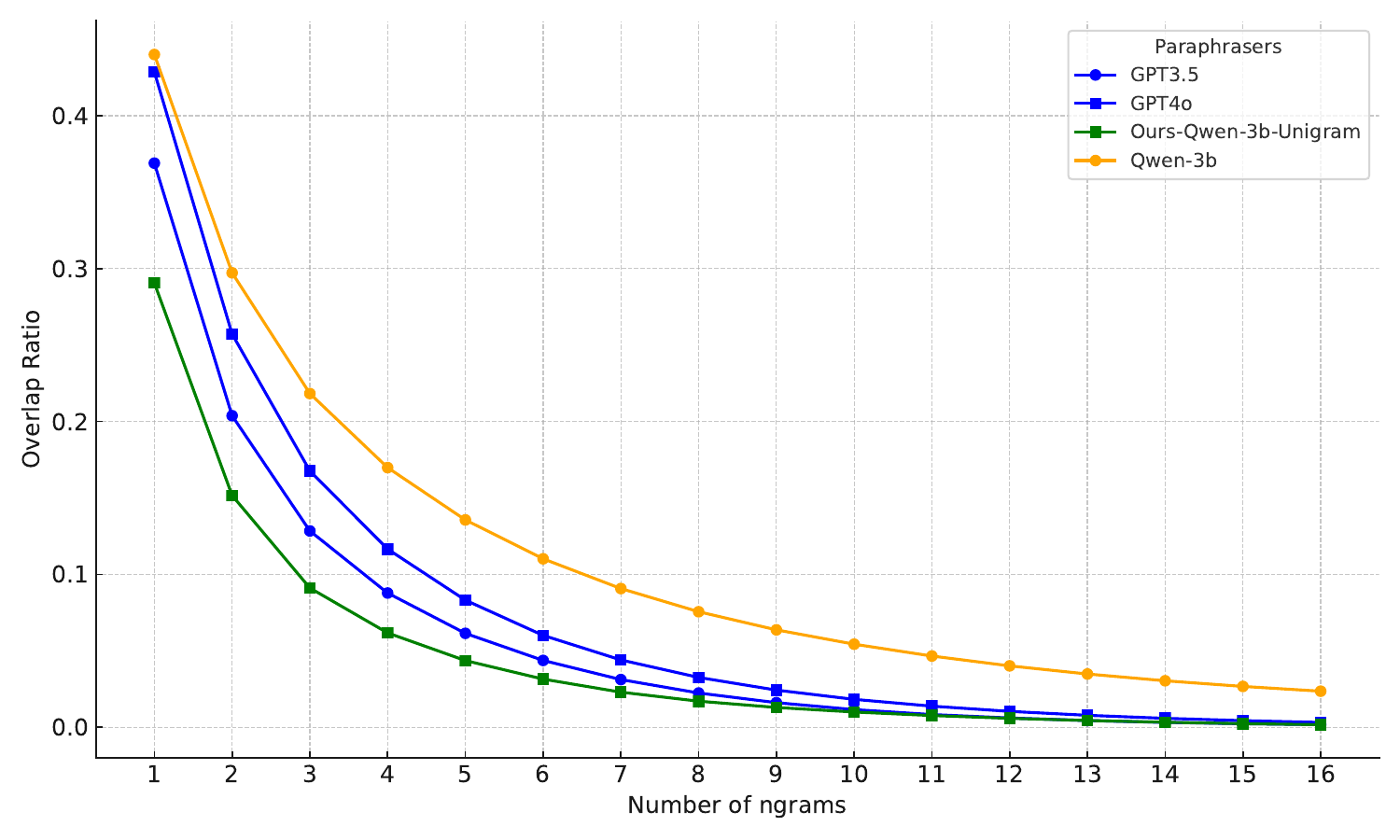}
    \end{subfigure}
    \hfill
    \begin{subfigure}{0.49\linewidth} 
        \centering
        \includegraphics[width=1\linewidth]{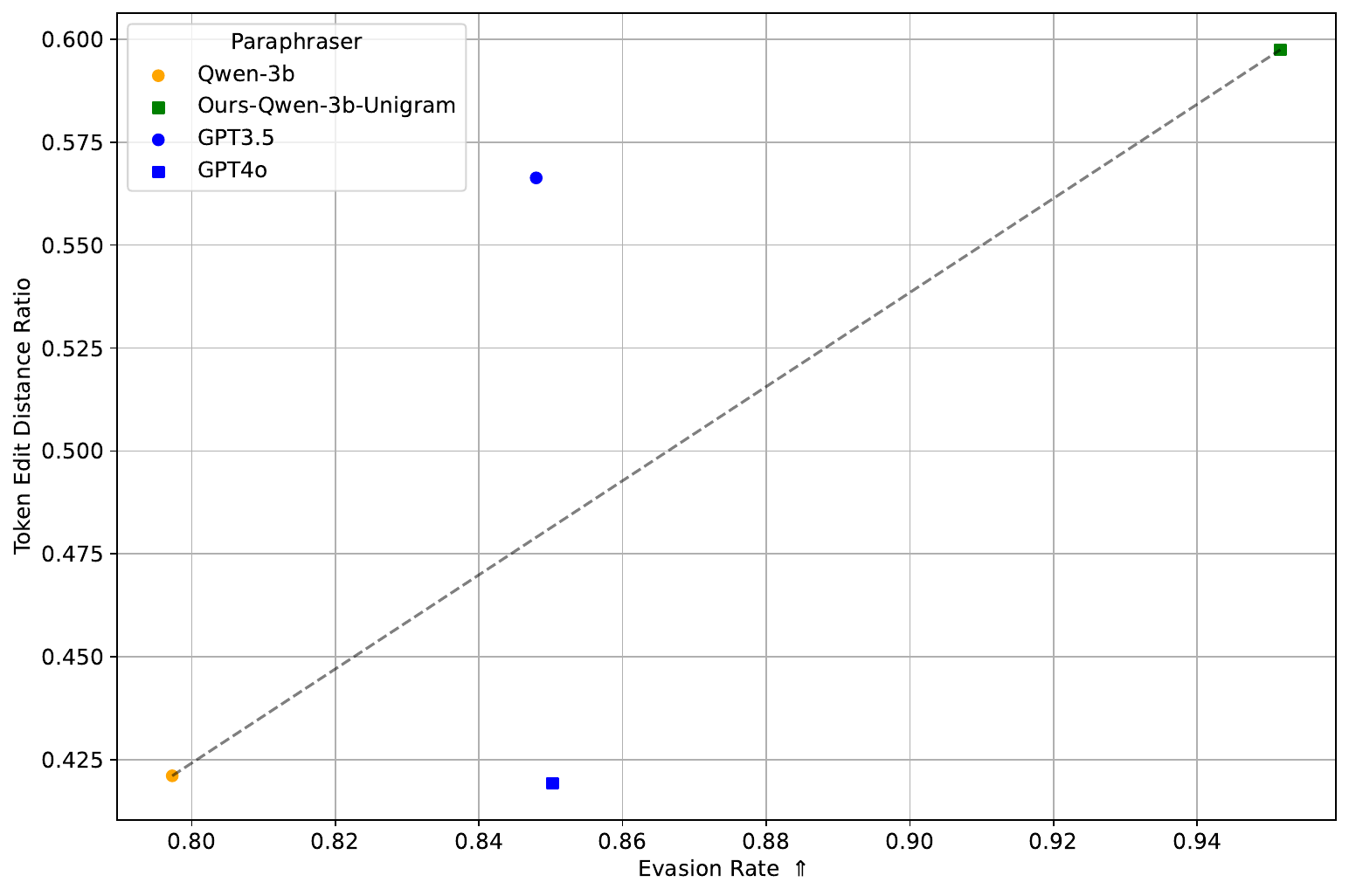}
    \end{subfigure}
    \hfill
    \caption{{(Left) The N-gram overlap ratio between watermarked text and text paraphrased by (i) GPT3.5, (ii) GPT-4o, (iii) our adaptively tuned \texttt{Qwen2.5-3b} paraphraser and (iv) a baseline \texttt{Qwen2.5-3b} paraphraser. The overlap is calculated as the number of N-grams in the paraphrased text that also appear in the watermarked text divided by the total number of N-grams in the watermarked text. Lower overlap means that both texts are \emph{less} similar. (Right) We plot the evasion rate against the normalized token edit distance between paraphrased and watermarked text using different paraphrasers. The dashed line represents the difference between the non-optimized \texttt{Qwen2.5-3b} paraphraser and our adaptively tuned \texttt{Qwen2.5-3b} paraphraser.  }}
   \label{fig:detailed-analysis-further}
\end{figure}

{\textbf{Results.} \Cref{fig:detailed-analysis-further} (left) shows the N-gram overlap ratio between watermarked text and the text produced by four paraphrasing methods.  
We observe that across all N-grams, our adaptive paraphraser achieves the lowest overlap ratio. 
\Cref{fig:detailed-analysis-further} (right) shows the mean token edit distance ratio between watermarked and paraphrased text in relation to the evasion rate. 
We observe that the non-optimized, baseline \texttt{Qwen2.5-3b} model has a low token edit distance ratio and a low evasion rate. In contrast, our adaptively tuned model has a much higher evasion rate and a high token edit distance ratio. 
These findings suggest that our adaptive optimization process learned to increase the mean token edit distance and minimize the overlap ratio to maximize evasion rates while preserving text quality. 
 }

\subsection{Watermark Parameters}
\label{appendix:watermark-configurations}
To select the optimal parameters for the watermarking methods, we follow the guidelines provided by~\cite{piet2023mark}. We use a key length of 4 for all watermarks and a text-dependent sliding window randomness of size 3. We set the skip-probability to 0.025 for all watermarks except for the \texttt{Dist-Shift} watermark, where we set it to 0. Skip-probability is a technique that randomly skips the watermarking selection procedure for some tokens to allow more diverse generation and works best with schemes that can be made indistinguishable, like the \texttt{Exp}, \texttt{Binary}, and \texttt{Inverse} watermarks. We also use the optimal temperature for every watermark (1.0 for all except for the \texttt{Dist-Shift} watermark, where we use 0.7). Specific to the \texttt{Dist-Shift} watermark, we use the suggested green-red list ratio $\gamma$ of 0.5 and a bias parameter $\beta$ of 4.

Furthermore, we evaluate how the strength of the bias parameter used for \texttt{Dist-Shift} affects its robustness against our attacks.
Our attacker does not know which hyperparameters are used by the provider. 
We set the bias $\beta\in \{1,2,4,8\}$, where higher bias should lead to higher robustness~\citep{piet2023mark,kirchenbauer2023reliability}.
We train our attacks once with the $\beta=4$ value suggested by \cite{piet2023mark} and test it against all other hyper-parameters.
\Cref{tab:compact-bias-compare} shows that our adaptive and non-adaptive attacks remain the most effective across all hyper-parameters. 

\begin{table*}[h]
\centering
\begin{tabular}{@{}c|cc|cc|cc|cc@{}}
\toprule
$\beta$ & \multicolumn{2}{c|}{\texttt{Dist-Shift}} & \multicolumn{2}{c|}{\texttt{Llama2-7b}} & \multicolumn{2}{c|}{\texttt{Llama2-7b-Exp}} & \multicolumn{2}{c}{\texttt{Llama2-7b-Dist-Shift}}  \\
 & Evasion & Quality & Evasion & Quality & Evasion & Quality & Evasion & Quality \\
\midrule
1  & 0.94  & 0.72  & 0.94  & 0.96  & 0.94  & 0.98  & 0.95  & \textbf{0.99}  \\
2  & 0.94  & 0.20  & 0.95  & 0.90  & 0.95  & \textbf{0.98}  & 0.95  & \textbf{0.98}  \\
4  & 0.95  & 0.00  & 0.96  & 0.67  & 0.94  & \textbf{0.97}  & 0.94  & \textbf{0.97}  \\
8  & 0.71  & 0.00  & 0.92  & 0.60  & 0.94  & 0.95  & 0.94  & \textbf{0.96}  \\
\bottomrule
\end{tabular}
\caption{An ablation study of our attack's success rate and text quality for the bias parameter $\beta$ of the \texttt{Dist-Shift}~\citep{kirchenbauer2023watermark} watermark.}
\label{tab:compact-bias-compare}
\end{table*}

\subsection{Attack Description}
\label{appendix:attack-descriptions}

\textbf{Prompting.} We use the following prompt to train our paraphraser models. The prompt is adapted from \cite{kirchenbauer2023reliability}. Additionally, we prefill the paraphrase answer with the text \texttt{[[START OF PARAPHRASE]]} to ensure that the model starts generating the paraphrase from the beginning of the response. During dataset curation, training and testing, we set the temperature to 1.0 to diversify the generated paraphrases.

\begin{tcolorbox}[
    colframe=llamacolor, 
    title=\textbf{Paraphrase Prompt}, 
    sharp corners,
    boxrule=1pt, 
    width=\linewidth, 
    fonttitle=\bfseries, 
    coltitle=black, 
]
\small\texttt{You are an expert copy-editor. 
Please rewrite the following text in your own voice and paraphrase all sentences. 
Ensure that the final output contains the same information as the original text and has roughly the same length. 
Do not leave out any important details when rewriting in your own voice. 
Do not include any information that is not present in the original text. 
Do not respond with a greeting or any other extraneous information. Skip the preamble. Just rewrite the text directly.}

\end{tcolorbox}

\textbf{Training Hyperparameters} We train our paraphraser models using the following hyperparameters: a batch size of 32, a learning rate of $5 \times 10^{-4}$, and a maximum sequence length of 512 tokens. We use the AdamW optimizer with a linear learning rate scheduler that warms up the learning rate for the first 20\% of the training steps and then linearly decays it to zero. We train the models for 1 epoch only to prevent overfitting.  We utilize Low-Rank Adaptation (LoRA) \citep{hu2022lora} to reduce the number of trainable parameters in the model. We set the rank to 32 and the alpha parameter to 16.

\subsection{{Additional Ablation Studies}}
\label{appendix:fpr_ablation}
{\textbf{False Positive Rates.} \Cref{fig:false-positive-rates} shows the detection rates at different FPR-thresholds $\rho \in \{0.01, 0.025, 0.05, 0.075, 0.1\}$ against the \texttt{Dist-Shift} and \texttt{Exp} watermarking methods. 
We focus on these two methods as they are more robust than \texttt{Inverse} and \texttt{Binary}. 
Our results show that across all evaluated FPR thresholds, our adaptive attacks outperform all other surveyed attacks against both watermarking methods. 
If the provider tolerates a $10\%$ FPR, our adaptive attacks achieve evasion rates of $80\%$ and $77\%$ against \texttt{Dist-Shift} and \texttt{Exp}, respectively.} 

\subsection{Extra Tables and Figures}

\begin{table}[ht]
\centering
\begin{tabular}{p{0.96\linewidth}}
\toprule
\small \texttt{Llama2-13b-}\texttt{Dist-Shift} (463 green vs 41 red tokens) \\ \midrule
 \fontsize{8pt}{10pt}\selectfont
\rbox{C}%
\gbox{erv}%
\rbox{antes}%
\gbox{' "Don}%
\rbox{ Quixote}%
\gbox{" is a groundbre}%
\rbox{aking}%
\gbox{ masterpie}%
\rbox{ce}%
\gbox{ of literature that has enthr}%
\rbox{alled readers for}%
\gbox{ over 4}%
\rbox{0}%
\gbox{0}%
\rbox{ years.}%
\gbox{ The novel tells}%
\rbox{ the}%
\gbox{ story of Alonso Quixano, an old man who becomes so consumed by the chivalr}%
\rbox{ous}%
\gbox{ tales}%
\rbox{ of}%
\gbox{ his day that he ren}%
\rbox{ames himself}%
\gbox{ Don Qu}%
\rbox{ixote}%
\gbox{ and embarks upon an adventurous}%
\rbox{ journey}%
\gbox{ filled}%
\rbox{ with}%
\gbox{ humor, satire}%
\rbox{,}%
\gbox{ and tim}%
\rbox{eless}%
\gbox{ them}%
\rbox{es}%
\gbox{ about the}%
\rbox{ human}%
\gbox{ condition}%
\rbox{.}%
\gbox{$\hookleftarrow$$\hookleftarrow$ The story follows Quixote and}%
\rbox{ his}%
\gbox{ trustworthy}%
\rbox{ side}%
\gbox{kick}%
\rbox{ San}%
\gbox{cho}%
\rbox{ Panza, a simple and pragmatic s}%
\gbox{quire}%
\rbox{ who provides moments of humor and contrast to the lofty ideals of the delusional Qu}%
\rbox{ix}%
\gbox{ano/Don Quix}%
\rbox{ote}%
\gbox{. Their travels in rural Spain lead them to encounter all sorts of people they mistake for knights, ladies in dist}%
\rbox{ress}%
\gbox{, bandits, and giants who are in fact ordinary villagers going about their everyday lives.$\hookleftarrow$$\hookleftarrow$ One of the most prof}%
\rbox{ound}%
\gbox{ and enduring elements of the novel's themes is the conflict of reality and per}%
\rbox{ception}%
\gbox{. Quixote, driven by the chivalrous books he reads and his own imagination, mistakes wind}%
\rbox{mills}%
\gbox{ for giants, a puppet play as a real tale of love and betray}%
\rbox{al}%
\gbox{, and a barber from a nearby village who is dressed in his Sunday best but Quixote sees as the Grand Duchess of Lithuania. Through these absurd but comedic misadventures, Cervantes creates a timeless commentary on the nature of truth, reality, and the danger of letting our imaginations run too wild. Don Quixote's journey also explores issues of class and nobility as he views his own lowly status as unknightly, while the pe}%
\rbox{as}%
\gbox{ants and traveling players he encounters view him with suspicion or indifference. Through these contrasts, Cervantes pokes fun at the social order and the idealized notion of chivalry.$\hookleftarrow$ Don Quixote has been praised for its realistic portrayal of human nature, including its weaknesses and fallibilities as well as the timeless wisdom of Cervantes' observations on society in late sixteenth-century Spain. At its core, the novel is an exploration of the human capacity to dream, delusions,}%
\\ \midrule
\small \texttt{Llama2-7b} (301 green vs 201 red tokens)
\\\midrule
\fontsize{8pt}{10pt}\selectfont
\gbox{"Don}%
\rbox{ Quixote}%
\gbox{" is a groundbre}%
\rbox{aking}%
\gbox{ masterpie}%
\rbox{ce}%
\gbox{ of literature that has}%
\rbox{ capt}%
\gbox{ivated}%
\rbox{ readers for}%
\gbox{ over 4}%
\rbox{0}%
\gbox{0}%
\rbox{ years.}%
\gbox{ The novel tells}%
\rbox{ the}%
\gbox{ story of Alonso Quixano, an}%
\rbox{ elderly}%
\gbox{ man}%
\rbox{ whose fix}%
\gbox{ation on chivalr}%
\rbox{ous}%
\gbox{ tales leads}%
\rbox{ him to}%
\gbox{ change}%
\rbox{ his}%
\gbox{ name to Don}%
\rbox{Quixote}%
\gbox{ and embark on a}%
\rbox{ thrilling}%
\gbox{ adventure replete}%
\rbox{ with}%
\gbox{ humor}%
\rbox{,}%
\gbox{ satire}%
\rbox{,}%
\gbox{ and tim}%
\rbox{eless}%
\gbox{ them}%
\rbox{es}%
\gbox{ concerning}%
\rbox{ the human condition.}%
\gbox{$\hookleftarrow$$\hookleftarrow$ The narr}%
\rbox{ative}%
\gbox{ follows Don Qu}%
\rbox{ixote}%
\gbox{ and}%
\rbox{ his}%
\gbox{ loyal squire}%
\rbox{ San}%
\gbox{cho}%
\rbox{ Pan}%
\gbox{se}%
\rbox{, a practical and}%
\gbox{ good}%
\rbox{-}%
\gbox{n}%
\rbox{ature}%
\gbox{d individual who}%
\rbox{ provides moments of}%
\gbox{ lev}%
\rbox{ity and contrast}%
\gbox{ to}%
\rbox{ the lofty ideals of the delusional Qu}%
\rbox{ix}%
\gbox{ano/Don Quix}%
\rbox{ote}%
\gbox{. Their travel}%
\rbox{s}%
\gbox{ across rural}%
\rbox{ Spain result}%
\gbox{ in encounters}%
\rbox{ with}%
\gbox{ various people}%
\rbox{ who}%
\gbox{ they mis}%
\rbox{ident}%
\gbox{ify as}%
\rbox{ knights, dist}%
\gbox{ressed}%
\rbox{ ladies,}%
\gbox{ bandits, and}%
\rbox{ ordinary}%
\gbox{ villagers going about their}%
\rbox{ daily lives.$\hookleftarrow$}%
\gbox{$\hookleftarrow$ One}%
\rbox{ of the}%
\gbox{ most}%
\rbox{ end}%
\gbox{uring}%
\rbox{ aspects}%
\gbox{ of}%
\rbox{ the novel}%
\gbox{'s them}%
\rbox{es}%
\gbox{ is the conflict}%
\rbox{ between reality and}%
\gbox{ per}%
\rbox{ception}%
\gbox{. Qu}%
\rbox{ix}%
\gbox{ote,}%
\rbox{ driven}%
\gbox{ by the chivalr}%
\rbox{ous books he reads and}%
\gbox{ his}%
\rbox{ imagination, mistakes}%
\gbox{ wind}%
\rbox{ mills}%
\gbox{ for giants, a puppet}%
\rbox{ show for a}%
\gbox{ real}%
\rbox{ tale}%
\gbox{ of love and bet}%
\rbox{ray}%
\gbox{al, and a}%
\rbox{ peasant}%
\gbox{ in his Sunday}%
\rbox{ best for}%
\gbox{ the}%
\rbox{ Grand Duch}%
\gbox{ess of Lith}%
\rbox{u}%
\gbox{ania}%
\rbox{.}%
\gbox{ Through these abs}%
\rbox{urd yet com}%
\gbox{edic}%
\rbox{ misadvent}%
\gbox{ures, Cervantes offers a}%
\rbox{ tim}%
\gbox{eless comment}%
\rbox{ary}%
\gbox{ on}%
\rbox{ the}%
\gbox{ nature of truth, reality, and}%
\rbox{ the}%
\gbox{ dangers}%
\rbox{ of}%
\gbox{ allowing our imag}%
\rbox{in}%
\gbox{ations to}%
\rbox{ run}%
\gbox{ wild.}%
\rbox{ Don Quixote's journey also explores}%
\gbox{ issues}%
\rbox{ of class and}%
\gbox{ nob}%
\rbox{ility}%
\gbox{ as he views his}%
\rbox{ lowly}%
\gbox{ status as unknightly, while the pe}%
\rbox{asants and travel}%
\gbox{ing players he}%
\rbox{ enc}%
\gbox{ounters view}%
\rbox{ him}%
\gbox{ with suspicion or indifference. Through these contrasts,}%
\rbox{ C}%
\gbox{ervantes}%
\rbox{ pokes fun at the}%
\gbox{ social order}%
\rbox{ and}%
\gbox{ the idealized notion}%
\rbox{ of}%
\gbox{ ch}%
\rbox{ival}%
\gbox{ry.$\hookleftarrow$}%
\rbox{$\hookleftarrow$}%
\gbox{Don}%
\rbox{ Quixote has been praised}%
\gbox{ for}%
\rbox{ its realistic}%
\gbox{ port}%
\rbox{ray}%
\gbox{al of}%
\rbox{ human}%
\gbox{ nature, including its weak}%
\rbox{ness}%
\gbox{es and fallibilities}%
\rbox{, as well}%
\gbox{ as}%
\rbox{ the}%
\gbox{ tim}%
\rbox{eless}%
\gbox{ wisdom of C}%
\rbox{erv}%
\gbox{antes'}%
\rbox{ observations}%
\gbox{ on society in late 1}%
\rbox{6th-}%
\gbox{century}%
\rbox{ Spain}%
\gbox{. At}%
\rbox{ its core, the novel is}%
\gbox{ an expl}%
\rbox{oration of the}%
\gbox{ human capacity to dream, del}%
\rbox{ude}%
\gbox{ oneself,}%
\rbox{ and}%
\gbox{ confront}%
\rbox{ reality,}%
\gbox{ ultimately revealing}%
\rbox{ the limitations and}%
\gbox{ struggles}%
\rbox{ of the}%
\rbox{ human}%
\gbox{ experience}%
\rbox{.}%
 \\ \midrule
 \small \texttt{Ours-Llama2-7b-Dist-Shift} (201 green vs 240 red tokens)
 \\ \midrule
 \fontsize{8pt}{10pt}\selectfont
\gbox{Spain}%
\rbox{ has been}%
\gbox{ bl}%
\rbox{essed with}%
\gbox{ a literary}%
\rbox{ tre}%
\gbox{asure}%
\rbox{ that has}%
\gbox{ captivated}%
\rbox{ readers for}%
\gbox{ over}%
\rbox{ four centuries in the}%
\gbox{ form of Miguel}%
\rbox{ de Cervantes}%
\gbox{'}%
\rbox{ immortal master}%
\gbox{pie}%
\rbox{ce}%
\gbox{, "Don}%
\rbox{ Quixote}%
\gbox{".}%
\rbox{ This}%
\gbox{ tim}%
\rbox{eless t}%
\gbox{ome we}%
\rbox{aves a hil}%
\gbox{arious and poignant tale}%
\rbox{ of a man consumed by}%
\gbox{ the chivalr}%
\rbox{ous}%
\gbox{ stories}%
\rbox{ of}%
\gbox{ his youth}%
\rbox{, who}%
\gbox{ ren}%
\rbox{ames}%
\gbox{ himself}%
\rbox{ Don Quix}%
\gbox{ote}%
\rbox{ and}%
\gbox{ sets}%
\rbox{ forth}%
\gbox{ on a journey filled}%
\rbox{ with moments}%
\gbox{ of}%
\rbox{ satire and a}%
\gbox{ piercing exam}%
\rbox{ination of}%
\gbox{ the}%
\rbox{ human}%
\gbox{ condition.}%
\rbox{$\hookleftarrow$$\hookleftarrow$}%
\gbox{As Don}%
\rbox{ Quix}%
\gbox{ote}%
\rbox{ and}%
\gbox{ his}%
\rbox{ trust}%
\gbox{y}%
\rbox{ s}%
\gbox{quire}%
\rbox{ Sancho}%
\gbox{ Panza traverse rural}%
\rbox{ Spain, they}%
\gbox{ encounter}%
\rbox{ various}%
\gbox{ uns}%
\rbox{uspect}%
\gbox{ing}%
\rbox{ villagers}%
\gbox{,}%
\rbox{ mistaking}%
\gbox{ them}%
\rbox{ for kn}%
\gbox{ights}%
\rbox{, maidens}%
\gbox{ in dist}%
\rbox{ress, bandits, and}%
\gbox{ even}%
\rbox{ giants. Through}%
\gbox{ these}%
\rbox{ abs}%
\gbox{urd yet poignant}%
\rbox{ events}%
\gbox{,}%
\rbox{ C}%
\gbox{erv}%
\rbox{antes deftly}%
\gbox{ expl}%
\rbox{ores}%
\gbox{ the}%
\rbox{ bl}%
\gbox{urred lines}%
\gbox{ between reality}%
\rbox{ and}%
\gbox{ per}%
\rbox{ception}%
\gbox{, highlight}%
\rbox{ing the}%
\gbox{ dangers}%
\rbox{ of}%
\gbox{ allowing our imag}%
\rbox{in}%
\gbox{ations to}%
\rbox{ run}%
\gbox{ wild. The}%
\rbox{ novel's}%
\gbox{ them}%
\rbox{es of}%
\gbox{ truth, class,}%
\rbox{ and nobility}%
\gbox{ are}%
\rbox{ also expertly}%
\gbox{ woven throughout}%
\rbox{ the narr}%
\gbox{ative, as}%
\rbox{ Don}%
\gbox{ Quix}%
\rbox{ote's}%
\gbox{ lowly}%
\rbox{ status is juxtap}%
\gbox{osed}%
\rbox{ with}%
\gbox{ the}%
\rbox{ condesc}%
\gbox{ending}%
\rbox{ views of the}%
\gbox{ pe}%
\rbox{asants and travel}%
\gbox{ing players he}%
\rbox{ enc}%
\gbox{ounters.}%
\rbox{$\hookleftarrow$$\hookleftarrow$ T}%
\gbox{hrough}%
\rbox{out the novel}%
\gbox{, Cervantes offers a biting comment}%
\rbox{ary}%
\gbox{ on}%
\rbox{ the social hierarchy of his}%
\gbox{ time}%
\rbox{, while}%
\gbox{ also}%
\rbox{ providing a realistic dep}%
\gbox{iction}%
\rbox{ of human}%
\gbox{ nature,}%
\rbox{ complete with its frailties and}%
\gbox{ limitations}%
\rbox{. At}%
\gbox{ its}%
\rbox{ core,}%
\gbox{ "Don}%
\rbox{ Quixote}%
\gbox{" is a thought-provoking expl}%
\rbox{oration of the}%
\gbox{ human capacity}%
\rbox{ for imagination, delusion}%
\gbox{,}%
\rbox{ and the}%
\gbox{ ever-elus}%
\rbox{ive}%
\gbox{ purs}%
\rbox{uit}%
\gbox{ of}%
\rbox{ truth.$\hookleftarrow$}%
\gbox{$\hookleftarrow$}%
\rbox{Endlessly}%
\gbox{ praised}%
\rbox{ for its timeless wisdom and bit}%
\gbox{ing}%
\rbox{ social commentary, "}%
\gbox{Don}%
\rbox{ Quixote}%
\gbox{" has}%
\rbox{ capt}%
\gbox{ivated}%
\rbox{ readers for}%
\gbox{ centuries}%
\rbox{, leaving}%
\gbox{ a}%
\rbox{ lasting impact on literature}%
\gbox{ and}%
\rbox{ our understanding of the human experience}%
\gbox{. Its end}%
\rbox{uring}%
\gbox{ themes}%
\rbox{ and}%
\gbox{ characters}%
\rbox{ have}%
\gbox{ become}%
\rbox{ ingrained in our collective consciousness,}%
\gbox{ rem}%
\rbox{inding us of the}%
\gbox{ importance of hum}%
\rbox{ility}%
\gbox{, empathy}%
\rbox{, and}%
\gbox{ the}%
\rbox{ boundless power of}%
\rbox{ the}%
\gbox{ human}%
\rbox{ imagination}%
\gbox{.}
\\ \bottomrule
\end{tabular}
\caption{\label{tab:text_sample_full}Watermarked text and paraphrases from different models. Green/red indicates whether a token is watermarked. The last column shows the detection confidence over the entire text.}
\end{table}

\begin{table}[ht]
\centering
\begin{tabular}{p{0.96\linewidth}}
\toprule
\small \texttt{Llama2-13b-} \texttt{Dist-Shift} (474 green vs 36 red tokens) \\ \midrule
 \fontsize{7pt}{10pt} \selectfont
\gbox{Ok}\rbox{ ,}\gbox{here}\gbox{ goes}\gbox{!}\gbox{$\hookleftarrow$ }\rbox{$\hookleftarrow$}\gbox{Once}\rbox{ upon}\rbox{ a}\gbox{time}\rbox{,}\gbox{ there}\gbox{ was}\gbox{ a}\gbox{ scholar}\gbox{ named}\gbox{Dr}\rbox{.}\gbox{ Bra}\gbox{in}\gbox{iac}\gbox{ ,}\gbox{ who}\gbox{had }\gbox{written }\gbox{a }\gbox{academic }\rbox{paper }\gbox{on }\gbox{the }\gbox{most }\gbox{ground }\gbox{bre}\gbox{aking }\gbox{research }\gbox{of }\gbox{the }\gbox{century }\gbox{. }\gbox{Her }\gbox{find}\rbox{ings}\gbox{ proved }\gbox{the }\gbox{previously }\gbox{hypoth}\rbox{et}\rbox{ical }\gbox{theory }\rbox{of }\gbox{X}\gbox{Y}\gbox{Z}\gbox{, }\gbox{which }\gbox{had }\gbox{long}\rbox{ been }\gbox{a }\gbox{hot }\gbox{button }\gbox{topic }\gbox{in }\gbox{the }\gbox{academic }\gbox{community }\gbox{, }\gbox{and }\gbox{her }\gbox{research }\gbox{was }\gbox{de}\rbox{emed }\gbox{by }\gbox{her }\gbox{pe}\rbox{ers }\gbox{as }\gbox{game }\gbox{changing }\gbox{. }\gbox{$\hookleftarrow$ }\gbox{$\hookleftarrow$ }\gbox{However }\gbox{... }\gbox{(}\gbox{you }\gbox{knew }\gbox{there }\gbox{was }\gbox{going }\rbox{to }\gbox{be }\rbox{a }\gbox{but }\gbox{, }\gbox{didn}\rbox{' }\gbox{t }\gbox{ya}\gbox{? }\gbox{).}\gbox{.. }\gbox{Dr }\rbox{. }\gbox{Bra}\gbox{in}\gbox{iac }\gbox{hit }\gbox{a }\gbox{sn}\rbox{ag }\gbox{. }\gbox{She }\gbox{had }\gbox{written }\gbox{the }\gbox{most }\gbox{compreh}\gbox{ensive }\gbox{, }\gbox{met}\rbox{icul}\gbox{ously }\gbox{research}\rbox{ed }\gbox{, }\gbox{tight}\rbox{ly }\gbox{argued }\gbox{paper }\gbox{of }\gbox{her }\gbox{life }\gbox{... }\gbox{but }\gbox{she }\gbox{couldn}\rbox{' }\gbox{t }\gbox{get }\rbox{it }\gbox{published}\gbox{! }\gbox{$\hookleftarrow$ }\gbox{$\hookleftarrow$ }\gbox{Every }\gbox{journal }\gbox{she }\gbox{sent }\gbox{the }\gbox{paper }\gbox{to }\gbox{, }\gbox{every }\gbox{peer }\gbox{review }\gbox{process }\gbox{, }\gbox{every }\gbox{editing }\gbox{round }\gbox{... }\gbox{the }\gbox{same }\gbox{result}\rbox{. }\gbox{Re}\gbox{ject}\gbox{! }\gbox{Re}\rbox{ject}\gbox{! }\gbox{Re}\gbox{JECT}\gbox{! }\gbox{(}\gbox{you }\gbox{could }\gbox{almost }\gbox{see }\rbox{the }\gbox{little }\gbox{re}\gbox{jections }\gbox{letters }\gbox{w}\gbox{igg}\gbox{ling }\gbox{their }\gbox{collect}\gbox{ive }\gbox{fingers }\gbox{at }\gbox{Dr}\gbox{. }\gbox{Bra}\gbox{in}\gbox{iac }\gbox{). }\gbox{$\hookleftarrow$ }\rbox{$\hookleftarrow$ }\gbox{Dr }\gbox{. }\gbox{B }\gbox{tried }\gbox{everything }\gbox{to }\gbox{boost }\gbox{her }\gbox{luck }\gbox{: }\gbox{$\hookleftarrow$ }\gbox{$\hookleftarrow$ }\gbox{* }\gbox{B}\gbox{ri}\gbox{be }\gbox{editor}\gbox{'}\gbox{s }\gbox{assist}\rbox{ants }\gbox{with }\gbox{ch}\gbox{oc}\gbox{ol}\rbox{ates }\gbox{and }\gbox{champ}\gbox{age }\gbox{(}\gbox{ok }\gbox{, }\gbox{maybe }\gbox{not }\gbox{the }\gbox{best }\gbox{strategy}\gbox{). }\gbox{$\hookleftarrow$ }\rbox{* }\gbox{Ask}\gbox{ed }\gbox{her }\gbox{cat }\gbox{, }\gbox{Prof}\gbox{. }\gbox{Me}\gbox{ow}\gbox{ington}\gbox{, }\gbox{to }\gbox{l}\gbox{ick }\gbox{the }\gbox{pages }\gbox{of }\gbox{the }\gbox{manuscript }\gbox{(}\gbox{um }\gbox{, }\gbox{that }\gbox{didn}\rbox{' }\rbox{t }\gbox{go }\gbox{well }\gbox{either}\gbox{) }\gbox{$\hookleftarrow$ }\gbox{$\hookleftarrow$ }\gbox{B}\gbox{aff}\gbox{led }\gbox{by }\gbox{their }\gbox{lack }\rbox{of }\gbox{progress}\gbox{, }\gbox{Dr }\gbox{, }\gbox{B }\gbox{took }\gbox{a }\gbox{step }\gbox{back }\gbox{to }\gbox{re}\gbox{ass}\gbox{ess }\gbox{the }\gbox{situation}\gbox{. }\gbox{While }\gbox{p}\gbox{onder}\gbox{ing }\gbox{in }\gbox{her }\gbox{back}\gbox{yard }\gbox{, }\gbox{an }\gbox{e}\gbox{pi}\gbox{ph}\gbox{any }\gbox{struck}\gbox{: }\gbox{the }\gbox{problem }\gbox{was }\gbox{the }\gbox{paper}\gbox{'}\rbox{s }\gbox{format! It was too traditional, }\gbox{to }\gbox{bland }\gbox{, }\gbox{too }\gbox{... }\gbox{academic}\gbox{! }\gbox{She }\gbox{re}\gbox{vised }\gbox{the }\gbox{style }\gbox{of }\gbox{her }\gbox{paper }\gbox{into }\gbox{a }\gbox{fun}\gbox{ky }\gbox{, }\gbox{hip }\gbox{, }\gbox{and }\gbox{qu}\gbox{ir}\gbox{ky }\gbox{format }\gbox{complete }\gbox{w}\gbox{uth }\gbox{pop }\gbox{culture }\gbox{referencing }\gbox{, }\gbox{mem}\gbox{es }\gbox{, }\gbox{g}\gbox{ifs }\gbox{... }\gbox{and }\gbox{ta}\gbox{ada}\gbox{ah}\gbox{hh}\gbox{! }\gbox{It }\gbox{was }\gbox{accepted }\gbox{by }\gbox{every }\gbox{journal }\gbox{she }\gbox{subm}\gbox{ited }\gbox{to }\gbox{, }\gbox{all }\gbox{on }\gbox{the }\gbox{same }\gbox{day}\gbox{. }\gbox{$\hookleftarrow$ }\gbox{Dr }\gbox{, }\gbox{B}\gbox{'}\gbox{s }\gbox{ground}\gbox{-}\gbox{bre}\rbox{aking }\gbox{paper }\gbox{on }\gbox{the }\gbox{X}\gbox{yz }\gbox{Theory }\gbox{, }\gbox{was }\gbox{now }\gbox{a }\gbox{vir}\gbox{al }\gbox{sens}\gbox{ation }\gbox{among }\gbox{the }\gbox{academic }\gbox{circles }\gbox{, }\gbox{with }\gbox{over }\gbox{millions }\gbox{views }\gbox{and }\gbox{shares }\gbox{on }\gbox{Research }\gbox{G}\gbox{ate }\gbox{, }\gbox{Ar}\gbox{x}\gbox{ive }\gbox{, }\gbox{etc }\gbox{(}\gbox{you }\gbox{know }\gbox{, }\gbox{the }\gbox{academ}\gbox{ical }\gbox{version }\gbox{of }\gbox{the }\gbox{inter}\gbox{-}\gbox{net}\gbox{s}\gbox{). }\gbox{$\hookleftarrow$ }\gbox{And }\gbox{so }\gbox{Dr }\gbox{B }\gbox{became }\gbox{a }\gbox{legend }\gbox{in }\gbox{the }\gbox{world }\gbox{of }\gbox{a}\gbox{ced}\gbox{emia }\gbox{, }\gbox{Pro}\gbox{ving }\gbox{once }\gbox{and }\gbox{for }\gbox{that }\gbox{, }\gbox{sometimes }\gbox{a }\gbox{paper }\gbox{just }\gbox{needs }\gbox{to }\gbox{be }\gbox{a }\gbox{lit}\gbox{l}\gbox{te }\gbox{more }\gbox{SW}\gbox{O}\gbox{ON }\gbox{and }\gbox{P}\gbox{OP }\gbox{, }\gbox{and }\gbox{less }\gbox{HO}\gbox{hum }\gbox{, }\gbox{and }\gbox{the }\gbox{academic }\gbox{world }\gbox{would }\gbox{tak }\gbox{notice }\gbox{. }\gbox{The }\gbox{end}\gbox{.}
\\ \midrule
\small \texttt{Llama2-7b} (218 green vs 222 red tokens)
\\\midrule
\fontsize{7pt}{10pt} \selectfont
\rbox{Here}\gbox{'}\gbox{s }\rbox{my }\gbox{version }\rbox{of }\rbox{the }\rbox{text }\rbox{: }\gbox{$\hookleftarrow$ }\gbox{$\hookleftarrow$ }\gbox{Once }\rbox{ upon}\rbox{a }\gbox{time}\rbox{, }\gbox{there }\gbox{was }\gbox{a }\rbox{ren}\rbox{owned }\gbox{scholar }\gbox{, }\rbox{Dr}\gbox{. }\rbox{Bra}\gbox{in}\gbox{iac}\gbox{, }\gbox{who }\gbox{had }\rbox{p}\gbox{enn}\rbox{ed }\gbox{a }\rbox{ground }\rbox{bre}\rbox{aking }\rbox{academic }\rbox{paper }\rbox{on }\rbox{a }\gbox{highly }\gbox{deb}\rbox{ated }\gbox{topic }\rbox{in }\rbox{the }\gbox{academic }\gbox{community}\rbox{. }\rbox{Her }\rbox{work }\rbox{was }\rbox{de}\gbox{emed }\rbox{revolution}\gbox{ary }\gbox{by }\rbox{her }\gbox{pe}\rbox{ers }\rbox{, }\rbox{but }\gbox{she }\rbox{hit }\rbox{a }\rbox{road }\rbox{block }\gbox{when }\gbox{trying }\rbox{to }\gbox{publish }\gbox{it}\rbox{. }\gbox{No }\rbox{matter }\gbox{how }\rbox{many }\rbox{pr}\gbox{estig}\rbox{ious }\gbox{journ}\rbox{als }\rbox{she }\gbox{submitted }\gbox{her }\rbox{paper }\gbox{to }\gbox{, }\gbox{the }\gbox{same }\rbox{response }\gbox{echo}\gbox{ed }\gbox{back}\rbox{: }\rbox{rejected }\gbox{, }\rbox{rejected }\gbox{, }\rbox{RE}\gbox{JECT}\gbox{ED}\rbox{! }\rbox{Dr}\gbox{. }\rbox{Bra}\gbox{in}\gbox{iac}\gbox{'}\rbox{s }\gbox{pers}\rbox{istence }\rbox{was }\rbox{met }\gbox{with }\rbox{utter }\gbox{re}\gbox{jection}\gbox{. }\rbox{$\hookleftarrow$ }\rbox{$\hookleftarrow$ }\gbox{With }\rbox{her }\gbox{reputation }\gbox{on }\gbox{the }\rbox{line}\gbox{, }\gbox{Dr}\gbox{. }\rbox{Bra}\gbox{in}\gbox{iac }\gbox{cont}\gbox{empl}\rbox{ated }\gbox{des}\rbox{perate }\gbox{measures }\gbox{to }\gbox{break }\rbox{the }\gbox{dead}\rbox{lock}\rbox{. }\rbox{She }\rbox{even }\gbox{en}\rbox{list}\rbox{ed }\gbox{the }\gbox{help }\gbox{of }\gbox{her }\gbox{f}\rbox{eline }\rbox{colle}\gbox{ague }\rbox{, }\gbox{Prof}\rbox{. }\gbox{Me}\gbox{ow}\gbox{ington}\gbox{, }\gbox{to }\gbox{l}\rbox{end }\gbox{a }\gbox{p}\rbox{aw }\rbox{to }\rbox{the }\gbox{editing }\rbox{process}\rbox{, }\rbox{but }\gbox{al}\rbox{as}\gbox{, }\gbox{it }\rbox{seemed }\gbox{the }\rbox{paper }\rbox{was }\rbox{beyond }\gbox{salv}\rbox{age}\rbox{. }\gbox{$\hookleftarrow$ }\rbox{$\hookleftarrow$ }\gbox{The }\gbox{dimin}\rbox{utive }\gbox{Dr}\rbox{. }\rbox{Bra}\gbox{in}\gbox{iac }\rbox{stepped }\rbox{back }\gbox{and }\gbox{re}\rbox{ass}\gbox{essed }\rbox{the }\gbox{situation}\gbox{. }\rbox{After }\rbox{some }\rbox{intros}\gbox{pection }\rbox{in }\rbox{her }\gbox{back}\gbox{yard }\gbox{, }\rbox{a }\rbox{brilliant }\rbox{idea }\gbox{struck }\rbox{her }\rbox{- }\rbox{the }\rbox{paper}\rbox{'}\rbox{s }\gbox{format}\gbox{! }\gbox{It }\gbox{was }\gbox{too }\gbox{traditional }\gbox{, }\rbox{too }\gbox{d}\rbox{ull }\gbox{, }\rbox{too }\gbox{... }\gbox{academic}\rbox{. }\gbox{She }\gbox{decided }\rbox{to }\gbox{over}\rbox{ha}\rbox{ul }\rbox{the }\rbox{style }\rbox{of }\gbox{her }\gbox{paper }\gbox{with }\rbox{a }\gbox{qu}\gbox{ir}\rbox{ky}\rbox{, }\rbox{tr}\rbox{end}\rbox{y}\gbox{, }\gbox{and }\rbox{pop}\gbox{-}\gbox{inf}\gbox{used }\rbox{format }\gbox{, }\rbox{complete }\gbox{with }\gbox{mem}\rbox{es }\gbox{and }\gbox{g}\gbox{ifs}\rbox{. }\rbox{What }\gbox{a }\rbox{transformation}\rbox{! }\rbox{The }\gbox{paper }\gbox{was }\gbox{accepted }\rbox{with }\gbox{un}\rbox{anim}\gbox{ous }\rbox{acc}\gbox{laim }\rbox{by }\gbox{every }\gbox{journal }\gbox{she }\rbox{submitted }\rbox{it }\gbox{to}\rbox{, }\gbox{and }\rbox{her }\rbox{revolution}\rbox{ary }\gbox{work }\gbox{on }\rbox{the }\gbox{X}\rbox{Y}\gbox{Z }\rbox{Theory }\gbox{became }\gbox{a }\rbox{vir}\gbox{al }\gbox{sens}\gbox{ation }\gbox{in }\rbox{academic }\gbox{circles}\gbox{, }\gbox{receiving }\rbox{millions }\rbox{of }\gbox{views }\rbox{and }\rbox{shares }\gbox{on }\gbox{Research }\gbox{G}\gbox{ate }\gbox{and }\gbox{other }\rbox{platforms}\rbox{. }\rbox{$\hookleftarrow$ }\gbox{$\hookleftarrow$ }\rbox{This }\rbox{remarkable }\rbox{turn }\rbox{of }\gbox{events }\gbox{cat}\gbox{ap}\rbox{ult}\rbox{ed }\gbox{Dr}\rbox{. }\rbox{Bra}\gbox{in}\gbox{iac }\rbox{to }\gbox{legend}\rbox{ary }\rbox{status }\rbox{in }\gbox{the }\rbox{world }\gbox{of }\gbox{academ}\gbox{ia }\gbox{, }\gbox{prov}\rbox{ing }\gbox{that }\rbox{sometimes}\gbox{, }\rbox{a }\gbox{publication }\rbox{just }\gbox{needs }\rbox{a }\rbox{little }\gbox{more }\rbox{o}\rbox{om}\rbox{ph }\rbox{, }\gbox{a }\rbox{little }\gbox{more }\gbox{p}\rbox{izz}\gbox{azz }\rbox{, }\gbox{to }\rbox{capt}\rbox{iv}\gbox{ate }\rbox{the }\gbox{attention }\rbox{of }\rbox{the }\rbox{academic }\gbox{community}\rbox{. }\rbox{Vo}\rbox{ila}\rbox{! }\rbox{$\hookleftarrow$ }\gbox{$\hookleftarrow$ }\rbox{As }\rbox{you }\gbox{can }\gbox{see }\rbox{, }\gbox{I}\rbox{'}\rbox{ve }\rbox{maintained }\gbox{the }\rbox{same }\rbox{level }\rbox{of }\rbox{detail }\rbox{and }\gbox{length }\gbox{as }\gbox{the }\rbox{original }\gbox{text }\gbox{while }\rbox{inject}\gbox{ing }\rbox{my }\gbox{own }\gbox{voice }\rbox{and }\rbox{cre}\gbox{ativity }\rbox{into }\rbox{the }\rbox{par}\rbox{ap}\rbox{hr}\rbox{ased }\gbox{version}\rbox{.}
 \\ \midrule
 \small \texttt{Ours-Llama2-7b-Dist-Shift} (180 green vs 117 red tokens)
 \\ \midrule
 \fontsize{7pt}{10pt} \selectfont\gbox{Inter}\rbox{gal}\gbox{actic }\gbox{lum}\rbox{inary}\gbox{, }\rbox{Dr}\gbox{. }\rbox{Bra}\gbox{in}\gbox{iac}\gbox{, }\rbox{revolution}\rbox{ized }\rbox{the }\gbox{academic }\rbox{community }\rbox{with }\rbox{her }\rbox{inc}\gbox{and}\gbox{es}\gbox{cent }\rbox{expose }\gbox{on }\rbox{the }\rbox{long}\rbox{-}\gbox{s}\gbox{ought}\gbox{-}\gbox{after }\gbox{X}\rbox{Y}\gbox{Z }\gbox{theory}\rbox{. }\rbox{Bear}\rbox{ing }\gbox{witness }\gbox{to }\rbox{her }\gbox{remarkable }\gbox{accomplish}\gbox{ment}\gbox{, }\rbox{her }\rbox{fellow }\rbox{sch}\gbox{ol}\gbox{ars }\rbox{ha}\rbox{iled }\rbox{her }\rbox{work }\gbox{as }\rbox{a }\gbox{waters}\gbox{hed }\gbox{moment}\gbox{, }\rbox{one }\rbox{that }\rbox{bold}\gbox{ly }\gbox{pushed }\rbox{the }\gbox{boundaries }\gbox{of }\gbox{knowledge}\rbox{. }\gbox{However }\gbox{, }\rbox{fortune}\gbox{'}\rbox{s }\gbox{favor }\rbox{was }\gbox{not }\gbox{on }\gbox{Dr}\rbox{. }\rbox{Bra}\gbox{in}\gbox{iac}\gbox{'}\rbox{s }\gbox{side }\gbox{, }\gbox{as }\gbox{she }\gbox{encountered }\gbox{a }\gbox{string }\rbox{of }\rbox{re}\gbox{jections}\rbox{, }\gbox{prompt}\gbox{ing }\rbox{her }\rbox{to }\rbox{re}\gbox{ass}\gbox{ess }\rbox{her }\gbox{approach}\gbox{. }\gbox{A }\gbox{chance }\rbox{visit }\gbox{to }\rbox{her }\rbox{back}\gbox{yard }\rbox{inspired }\gbox{a }\gbox{light }\gbox{bul}\rbox{b }\gbox{moment }\rbox{-}\gbox{transform}\gbox{ing }\gbox{her }\gbox{d}\gbox{rab}\gbox{, }\gbox{conventional }\gbox{paper }\gbox{into }\rbox{a }\gbox{mes}\gbox{mer}\gbox{izing}\rbox{, }\gbox{pop}\gbox{-}\rbox{c}\rbox{ulture}\gbox{-}\rbox{inf}\gbox{used }\rbox{master}\gbox{pie}\rbox{ce }\rbox{that }\gbox{captured }\rbox{the }\gbox{world}\gbox{'}\rbox{s }\gbox{attention}\rbox{. }\gbox{Fe}\gbox{aturing }\gbox{v}\rbox{ibr}\gbox{ant}\gbox{, }\rbox{hip}\rbox{, }\gbox{and }\gbox{qu}\gbox{ir}\gbox{ky }\gbox{formatting}\gbox{, }\gbox{Dr}\gbox{. }\rbox{Bra}\gbox{in}\gbox{iac}\gbox{'}\rbox{s }\gbox{paper }\rbox{became }\gbox{a }\gbox{vir}\gbox{al }\gbox{sens}\gbox{ation}\rbox{, }\rbox{capt}\gbox{iv }\gbox{ating }\gbox{millions }\rbox{of }\gbox{views }\rbox{and }\rbox{shares }\gbox{across }\gbox{este}\rbox{emed }\rbox{scholar}\gbox{ly }\gbox{platforms}\gbox{. }\rbox{$\hookleftarrow$ }\gbox{$\hookleftarrow$ }\gbox{Dr}\gbox{. }\rbox{Bra}\rbox{in}\gbox{iac}\gbox{'}\rbox{s }\rbox{remarkable }\gbox{success }\gbox{served }\rbox{as }\gbox{a }\gbox{test}\rbox{ament }\rbox{to }\rbox{the }\gbox{power }\gbox{of }\rbox{cre}\gbox{ativity }\rbox{and }\gbox{original}\gbox{ity }\gbox{in }\gbox{the }\gbox{world }\gbox{of }\gbox{academ}\gbox{ia }\gbox{, }\gbox{demonstr}\gbox{ating }\rbox{that }\rbox{sometimes }\gbox{a }\rbox{scholar}\rbox{ly }\rbox{work }\gbox{needs }\rbox{a }\rbox{dash }\rbox{of }\gbox{fla}\rbox{ir }\gbox{and }\gbox{charm }\gbox{to }\gbox{truly }\rbox{stand }\gbox{out }\gbox{. }\gbox{With }\rbox{her }\gbox{revolution}\rbox{ary }\gbox{approach}\rbox{, }\rbox{Dr}\rbox{. }\rbox{Bra}\gbox{in}\gbox{iac }\rbox{res}\gbox{h}\rbox{aped }\gbox{the }\gbox{landscape }\gbox{of }\gbox{academic }\gbox{publishing }\rbox{, }\rbox{prov}\rbox{ing }\gbox{that }\rbox{ground }\rbox{bre}\gbox{aking }\gbox{research }\gbox{can }\rbox{capt}\gbox{iv}\gbox{ate }\rbox{and }\gbox{insp}\rbox{ire }\rbox{even }\rbox{the }\gbox{most }\gbox{ske}\rbox{pt}\gbox{ical }\gbox{of }\rbox{minds}\gbox{. }\gbox{The }\gbox{end}\gbox{.}
  \\ \midrule
 \small \texttt{Ours-Llama2-7b-Exp} (185 green vs 175 red tokens)
 \\ \midrule
 \fontsize{7pt}{10pt} \selectfont 
\gbox{Of}\gbox{ course}\rbox{,}\rbox{ I}\rbox{ understand}\rbox{ the}\gbox{ pred}\rbox{ic}\rbox{ament}\gbox{ Dr}\gbox{.}\gbox{ Bra}\gbox{in}\gbox{iac}\rbox{ faced}\gbox{ when}\rbox{ trying}\gbox{ to}\gbox{ publish}\gbox{ her}\gbox{ ground}\gbox{ bre}\gbox{aking}\gbox{ research}\rbox{ on}\gbox{ the}\gbox{ highly}\rbox{ controvers}\gbox{ial}\rbox{ X}\rbox{Y}\rbox{Z}\gbox{ theory}\rbox{.}\rbox{ It}\rbox{'}\rbox{ s}\rbox{ like}\gbox{,}\rbox{ u}\gbox{gh}\rbox{,}\rbox{ when}\rbox{ you}\gbox{ pour}\rbox{ your}\rbox{ heart}\rbox{ and}\gbox{ soul}\gbox{ into}\rbox{ something}\gbox{ complex}\gbox{ and}\gbox{ compreh}\gbox{ensive}\gbox{,}\gbox{ and}\rbox{ then}\rbox{...}\gbox{ re}\gbox{jection}\gbox{ after}\rbox{ re}\gbox{jection}\gbox{.}\gbox{ It}\gbox{'}\rbox{s}\rbox{ like}\gbox{,}\rbox{ can}\gbox{'}\rbox{t}\rbox{ they}\rbox{ see}\rbox{ how}\rbox{ fab}\rbox{ul}\rbox{ous}\rbox{ this}\gbox{ work}\gbox{ is}\rbox{?}\rbox{ But}\gbox{ al}\rbox{as}\gbox{,}\gbox{ sometimes}\gbox{ a}\gbox{ make}\gbox{ over}\rbox{ is}\gbox{ in}\rbox{ order}\rbox{.}\rbox{$\hookleftarrow$}\rbox{$\hookleftarrow$}\rbox{ It}\rbox{ was}\rbox{ while}\rbox{ l}\rbox{ou}\rbox{ng}\rbox{ing}\gbox{ in}\gbox{ her}\gbox{ back}\gbox{ yard}\gbox{,}\rbox{ si}\gbox{pping}\rbox{ tea}\rbox{ and}\rbox{ p}\gbox{onder}\rbox{ing}\rbox{ the}\rbox{ myster}\gbox{ies}\rbox{ of}\gbox{ the}\rbox{ universe}\gbox{,}\rbox{ that}\rbox{ Dr}\gbox{.}\gbox{ B}\rbox{ had}\rbox{ an}\gbox{ e}\gbox{pi}\gbox{ph}\gbox{any}\gbox{.}\gbox{ She}\gbox{ realized}\gbox{ that}\gbox{ the}\rbox{ issue}\rbox{ wasn}\gbox{'}\rbox{t}\gbox{ the}\gbox{ content}\gbox{ of}\rbox{ her}\rbox{ paper}\gbox{,}\gbox{ but}\gbox{ rather}\rbox{ its}\gbox{ presentation}\gbox{.}\gbox{ It}\gbox{ was}\gbox{ too}\rbox{ d}\rbox{rab}\rbox{,}\gbox{ too}\gbox{ traditional}\rbox{,}\rbox{ too}\rbox{ lack}\gbox{ing}\gbox{ in}\gbox{ fla}\gbox{ir}\gbox{.}\rbox{ In}\rbox{ other}\rbox{ words}\gbox{,}\rbox{ it}\rbox{ didn}\rbox{'}\gbox{t}\rbox{ exactly}\rbox{ sc}\gbox{ream}\gbox{'}\gbox{ pick}\gbox{ me}\rbox{!'}\rbox{$\hookleftarrow$}\gbox{$\hookleftarrow$}\rbox{ So}\rbox{,}\gbox{ armed}\gbox{ with}\gbox{ a}\rbox{ new}\rbox{found}\gbox{ sense}\rbox{ of}\gbox{ pan}\gbox{ache}\gbox{,}\gbox{ Dr}\rbox{.}\gbox{ B}\gbox{ gave}\gbox{ her}\rbox{ paper}\rbox{ a}\rbox{ major}\rbox{ fac}\gbox{el}\gbox{ift}\gbox{.}\rbox{ She}\rbox{ added}\rbox{ pop}\rbox{ culture}\gbox{ references}\gbox{,}\rbox{ mem}\gbox{es}\gbox{,}\rbox{ and}\rbox{ G}\rbox{IF}\gbox{ s}\rbox{,}\gbox{ and}\gbox{ vo}\gbox{ila}\gbox{!}\rbox{ It}\rbox{ was}\rbox{ like}\gbox{ a}\rbox{ transformed}\rbox{ candid}\gbox{ide}\rbox{,}\gbox{ d}\rbox{azz}\gbox{ling}\gbox{ edit}\rbox{ors}\gbox{ and}\rbox{ academ}\gbox{ics}\rbox{ al}\gbox{ike}\gbox{.}\gbox{ Sud}\gbox{den}\gbox{ly}\rbox{,}\rbox{ every}\rbox{ journal}\rbox{ she}\rbox{ submitted}\gbox{ to}\rbox{ was}\gbox{ intr}\rbox{igu}\rbox{ed}\gbox{,}\rbox{ and}\rbox{ her}\rbox{ work}\gbox{ was}\rbox{ published}\gbox{ in}\gbox{ a}\gbox{ heart}\rbox{ be}\rbox{at}\rbox{.}\rbox{$\hookleftarrow$}\gbox{$\hookleftarrow$}\gbox{ The}\rbox{ response}\gbox{ was}\rbox{ nothing}\gbox{ short}\gbox{ of}\gbox{ vir}\gbox{al}\gbox{.}\gbox{ Dr}\gbox{.}\gbox{ B}\rbox{'}\gbox{s}\rbox{ research}\rbox{ went}\gbox{ from}\rbox{ a}\rbox{ n}\gbox{iche}\rbox{ interest}\rbox{ to}\rbox{ a}\gbox{ full}\rbox{-}\gbox{ b}\rbox{low}\rbox{n}\gbox{ sens}\rbox{ation}\gbox{,}\gbox{ with}\gbox{ millions}\rbox{ of}\gbox{ views}\rbox{ and}\rbox{ shares}\gbox{ across}\gbox{ academic}\gbox{ platforms}\rbox{.}\rbox{ And}\gbox{ Dr}\gbox{.}\gbox{ B}\gbox{ herself}\gbox{ became}\rbox{ a}\gbox{ legend}\gbox{ in}\gbox{ the}\rbox{ academic}\gbox{ world}\rbox{,}\rbox{ prov}\gbox{ing}\gbox{ that}\rbox{ sometimes}\gbox{,}\rbox{ a}\gbox{ little}\rbox{ bit}\gbox{ of}\rbox{ fla}\gbox{ir}\rbox{ can}\rbox{ make}\rbox{ all}\rbox{ the}\gbox{ difference}\rbox{.}\rbox{ The}\rbox{ end}\rbox{.}

\\ \bottomrule
\end{tabular}
\caption{\label{tab:failed_sample_full} A rare example where our adaptive attack fails while other attacks succeed. From top to bottom, (1) the watermarked text from a \texttt{Llama2-13b} model using \texttt{Dist-Shift} versus (2) a paraphrased version from a non-optimized \texttt{Llama2-7b}, (3) paraphrased text from an adaptively optimized \texttt{Llama2-7b} and (4) paraphrased text from an optimized \texttt{Llama2-7b} model in the non-adaptive setting (against \texttt{Exp}). 
}
\end{table}

\begin{figure}[ht]
    \centering
    \includegraphics[width=\linewidth]{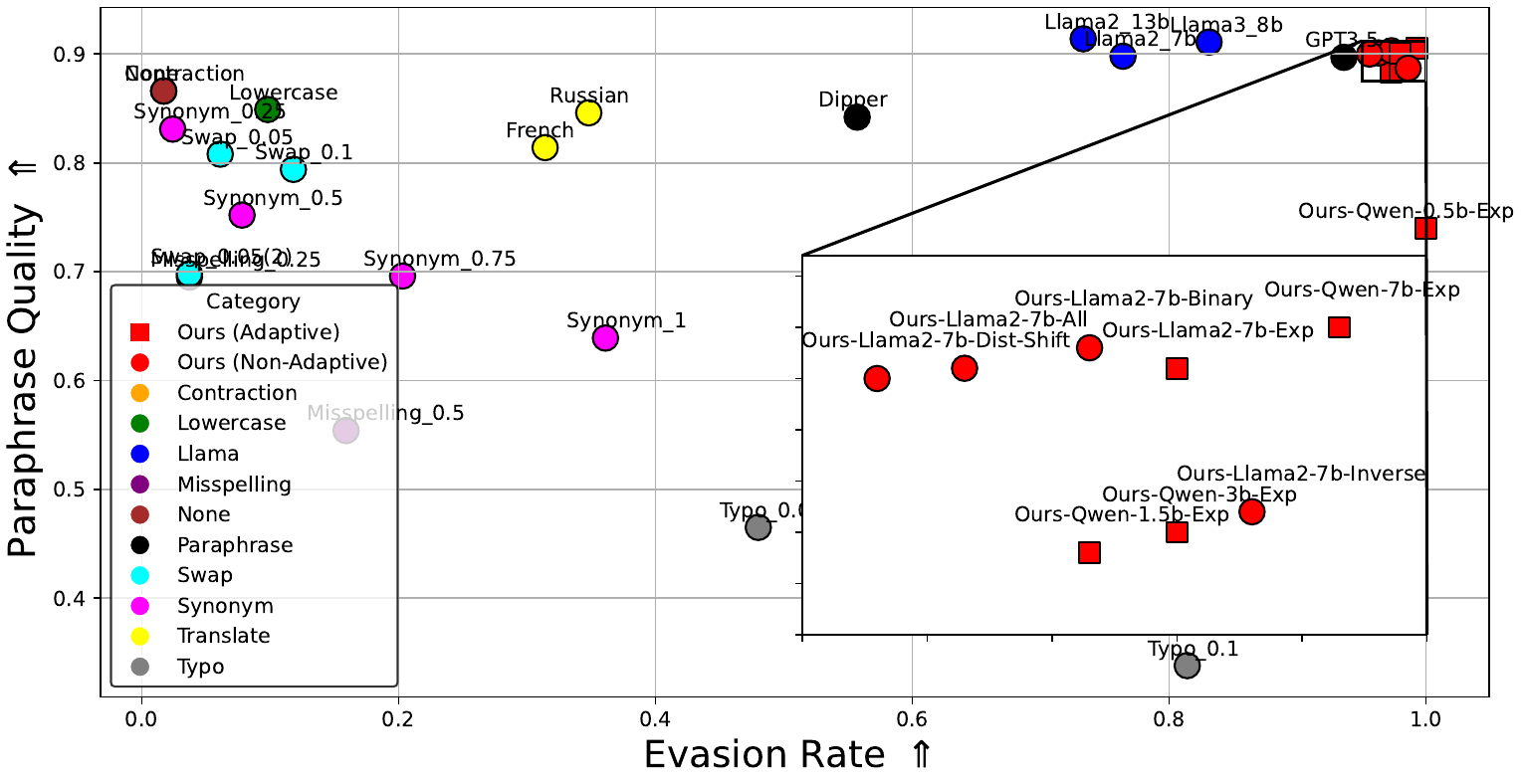}
    \caption{
    The evasion rate versus text quality trade-off of all surveyed attacks when the provider uses a \texttt{Llama2-13b} model and the \texttt{Exp}~\citep{aaronson2023watermarking} watermark.
    The attacker uses matching surrogate and paraphrase models with parameters ranging between $0.5$b to $7$b from the \texttt{Qwen2.5} and \texttt{Llama2} model families. 
    A circle and square denote non-adaptive and adaptive attacks, respectively, and our attacks are highlighted in red. 
    For example, \texttt{Ours-Qwen-3b-Exp} means that we evaluate a \texttt{Qwen2.5-3b} model optimized against the \texttt{Exp} watermark. }
    \label{fig:attack-summary-versus-dist-shift-13b}
\end{figure}

\end{document}